# Probing C–I Bond Fission in the UV Photochemistry of 2-Iodothiophene with Core-to-Valence Transient Absorption Spectroscopy

Benjamin W. Toulson,[1,*] Diptarka Hait,[1,2,a,*] Davide Faccialà,[3] Daniel M. Neumark,[1,2] Stephen R. Leone,[1,2,4] Martin Head-Gordon,[1,2] and Oliver Gessner[1,⊥]

[1]Chemical Sciences Division, Lawrence Berkeley National Laboratory, Berkeley, California 94720, USA
[2]Department of Chemistry, University of California, Berkeley, California 94720, USA
[3]CNR-Istituto di Fotonica e Nanotecnologie (CNR-IFN), 20133 Milano, Italy
[4]Department of Physics, University of California, Berkeley, California 94720, USA
[a]Current Address: Department of Chemistry and PULSE Institute, Stanford University, Stanford, CA 94305, USA

[*] These authors contributed equally to this work.
[⊥] email: ogessner@lbl.gov

# Abstract

The UV photochemistry of small heteroaromatic molecules serves as a testbed for understanding fundamental photo-induced chemical transformations in moderately complex compounds, including isomerization, ring-opening, and molecular dissociation. Here, a combined experimental-theoretical study of 268 nm UV light-induced dynamics in 2-iodothiophene ($C_4H_3IS$) is performed. The dynamics are experimentally monitored with a femtosecond XUV probe that measures iodine N-edge 4d core-to-valence transitions. Experiments are complemented by density functional theory calculations of both the pump-pulse induced valence excitations as well as the XUV probe-induced core-to-valence transitions. Possible intramolecular relaxation dynamics are investigated by *ab initio* molecular dynamics simulations. Gradual absorption changes up to ~0.5-1 ps after excitation are observed for both the parent molecular species and emerging iodine fragments, with the latter appearing with a characteristic rise time of 160±30 fs. Comparison of spectral intensities and energies with the calculations identify an iodine dissociation initiated by a predominant π→π* excitation. In contrast, initial excitation to a nearby $n_\perp$→σ* state appears unlikely based on a significantly smaller oscillator strength and the absence of any corresponding XUV absorption signatures. Excitation to the π→π* state is followed by contraction of the C-I bond, enabling a nonadiabatic transition to a dissociative π→$\sigma^*_{C-I}$ state. For the subsequent fragmentation, a relatively narrow bond-length region along the C-I stretch coordinate between 230 and 280 pm is identified, where the transition between the parent molecule and the thienyl radical + iodine atom products becomes prominent in the XUV spectrum due to rapid localization of two singly-occupied molecular orbitals on the two fragments.

# 1. Introduction

The UV photochemistry of isolated cyclic molecules provides an important testbed for understanding electronic structures and associated coupled electronic-nuclear dynamics in moderately complex molecular compounds. Of particular interest are nonadiabatic transitions between electronic states at conical intersections (CI), which are thought to be a common mechanism, for example, to provide an efficient route to non-radiative, non-destructive relaxation. [1]  An alternative fate is photodamage, where the surface crossing can potentially lead to states with antibonding $\sigma^*$ character, driving bond fission pathways such as ring-opening and photofragmentation. The experimental and theoretical understanding of photoinitiated heterocyclic ring-opening has been surveyed recently.[2]  Sulfur-containing compounds are gaining additional recognition, as they are ubiquitous in biological and renewable energy relevant systems, and their photochemistry is particularly rich.[3-6]

The first absorption band of halothiophenes comprises a broad absorption continuum centered near 240 nm with a full width at half maximum (FWHM) of ~40 nm and exhibits superimposed diffuse vibrational structure.[7-9]  In iodothiophene, vibronic resonances are prominent on the red-edge of the absorption band and separated by ~630 cm$^{-1}$.[7]  Previous *ab initio* calculations [7] suggest that two types of initial excitation are responsible for this band: diabatically bound $\pi \rightarrow \pi^*$ excitations that are responsible for the vibrational structure in the absorption spectrum, and dissociative $(n_I/\pi) \rightarrow \sigma^*_{C-I}$ excited states resulting in a broad absorption continuum.[7]  The notation $(n_I/\pi) \rightarrow \sigma^*_{C-I}$ denotes the promotion of an electron into an antibonding $\sigma^*$ molecular orbital between the carbon and iodine from a combination of a nonbonding *p* orbital on the iodine atom and from a $\pi$ orbital of the aromatic ring.

Following single-photon UV excitation of halothiophenes, several product channels are energetically accessible, including C–X bond fission as well as C–S ring-opening pathways to multiple isomers.[10]  The gas phase photochemistries of 2-iodothiophene and 2-bromothiophene have previously been studied across the first absorption band (λ = 220–303 nm) using velocity-map ion imaging.[7,11] Halogen atom photoproducts were state-selectively ionized in their ground and valence-excited electronic states. The trends of translational energy release into the photofragments with wavelength for both halothiophenes are similar. Large translational energies and pronounced fragment ejection anisotropies are observed at longer photolysis wavelengths and associated with dissociation on a repulsive surface. A second product channel becomes increasingly important at shorter photolysis wavelengths, characterized by smaller translational energies (and by conservation of energy, indicating the production of a highly internally excited $C_4H_3S$ co-fragment from dissociation) and an isotropic fragment emission.

Resonance Raman spectra of 2-iodothiophene (in solution) following ~240 nm - 253 nm excitation indicate strong activation of ring stretching modes; parallels to the photochemistry of iodobenzene are drawn as the resonance Raman spectra show similar ring excitation.[12] Iodobenzene is well studied both experimentally and theoretically, showing complex wavelength-dependent photodissociation dynamics.[13-18] The first absorption band comprises both dissociative states of (n/π)→ $\sigma^*_{C-I}$ character and also bound π→π* states that are coupled to dissociative n→ $\sigma^*_{C-I}$ and/or π→ $\sigma^*_{C-I}$ states.[18] Drescher et al. used XUV transient absorption spectroscopy at the iodine $N_{4,5}$ edge to study UV-induced dissociation after 268 nm excitation of iodobenzene and iodomethane.[19] While atomic iodine signals appeared within the ~100 fs resolution of the experiment for iodomethane, a delay of ~40 fs was observed for the corresponding signals in iodobenzene.[19] The different responses were attributed to different distances of the iodine reporter atom from the UV-induced vacancies in both cases, i.e., a $(2e)^{-1}$ hole localized on iodine in the case of iodomethane and a $\pi^{-1}$ vacancy localized on the phenyl ring for iodobenzene. It was speculated that the delay is due to valence electron dynamics required to migrate the hole closer to the I-atom and activate XUV transitions on the atomic fragment.[19] Attar et al. had previously used the same technique to study 266 nm induced dissociation in methyl iodide. The evolution of the XUV transient absorption signal over the first ~100 fs was assigned to a passage through a transition-state region after ~40 fs, and completion of the C-I dissociation within ~90 fs.[20] Chang et al. studied UV induced dissociation in a series of alkyl iodides of different sizes using broadband, ~20 fs pump pulses centered at 277 nm.[21] Using these short excitation pulses, the inverse rate constants (from fitting to an exponential) for the appearance of neutral iodine fragments were determined to vary between ~25 fs for methyl iodide and ~49 fs for *t*-$C_4H_9I$, and to systematically increase with increasing molecular size. Intriguingly, the time delay for the appearance of photoproduct signals can be sensitive to the experimental probe. Using femtosecond UV pump – multi-photon ionization probe time-of-flight mass spectrometry, Zewail and co-workers observed signatures of neutral iodine atoms appearing on timescales of ~90 fs for iodomethane[22] and ≥400 fs for iodobenzene[14,23] after 268 nm - 278 nm excitation. While the appearance time in iodomethane agrees with the findings of Attar et al.[20] in iodobenzene, the I atom appearance is significantly slower than observed by XUV transient absorption spectroscopy.[19] The various findings and interpretations of spectroscopic trends in UV-induced dissociation dynamics of small, iodine containing molecules illustrate the importance of further careful theoretical examination of the spectroscopic signatures and underlying coupled electronic-nuclear dynamics to translate experimental observables into photochemical insight.

Here, a UV pump – XUV probe femtosecond transient absorption spectroscopy study of the 268 nm dissociation of 2-iodothiophene is presented. The XUV spectra track the formation of singly occupied molecular orbitals (SOMO) in the photoexcitation step, the rapid appearance of ionic species from dissociative multi-photon ionization, and the evolution of transient electronic configurations throughout the C–I bond fission to produce neutral iodine fragments. Pump pulse intensity dependent XUV spectra indicate that neutral and cationic iodine fragments result from independent pathways, enabling a detailed analysis of the neutral dissociation mechanism despite the appearance of ionic signals. A global fit of the experimental XUV spectra is used to decompose the spectral and temporal trends, which are interpreted with the aid of *ab initio* calculations that predict the energies and oscillator strengths of both valence and inner-shell transitions of the pump- and probe-steps, respectively. For the neutral dissociation channel, the combination of experimental and theoretical findings identifies a π→π* configuration as the predominant initially excited state, after which the molecular wavepacket first evolves toward shorter C-I distances compared to the electronic ground state ($S_0$) equilibrium, and then the system transfers to a dissociative π→σ* potential energy surface via a conical intersection (henceforth we drop the C-I subscript for σ* orbitals). Participation of an energetically accessible $n_\perp$→σ* state, corresponding to excitation of a nonbonding iodine lone pair electron to an antibonding σ* orbital, is disfavored due to a much smaller oscillator strength (≤1/6 of π→π*) of the valence excitation and the lack of any XUV signature at the calculated $n_\perp$→σ* energy at short-time pump-probe delays. The characteristic timescale of the iodine elimination is experimentally determined to be ~160 fs, whereby the bulk of the key electronic structure rearrangements connecting the united molecule and the separated atom spectroscopic regimes occur within a narrow bond-length region of 230-280 pm C-I distance.

## 2. Experimental and Theoretical Methods

### 2.1. Femtosecond XUV Transient Absorption Spectroscopy

Experiments are performed using a femtosecond XUV transient absorption setup that has been described previously[24] and following the method outlined in our recent publication on UV-induced dynamics in $CHBr_3$.[25] In brief, high-harmonic generation (HHG) produces a quasi-continuum of XUV by focusing 4.0 mJ, 35 fs, 1 kHz, 804 nm NIR pulses with a 500 mm focal length lens into a semi-infinite gas cell containing 65 Torr of Argon. The NIR is then rejected by a 200 nm thick aluminum foil and the transmitted XUV light impinges on a toroidal mirror that focuses the beam into an ~4 mm long sample cell containing 2-iodothiophene (Sigma Aldrich, 98%). The transmitted XUV is detected by a spectrometer consisting of a variable line spaced grating (~1200 lines/mm) and an X-ray CCD camera. The sample chamber is isolated from the other vacuum chambers using 200 nm thick aluminum foils to prevent contamination of optics and the camera.

Transient absorption spectra are measured by introducing a pump pulse into the sample cell. In the pump arm, up to 4.0 mJ of NIR are used to generate UV pump pulses by frequency-tripling the fundamental to produce up to 300 μJ pulses of UV light at 268 nm. Experiments are performed at pump pulse intensities between ~1 TW/cm$^2$ and ~5 TW/cm$^2$. The polarization of both the UV and the HHG beams is horizontal in the laboratory frame. The UV beam is focused into the sample cell with a 500 mm focal length lens to a spot diameter of ~100 μm (FWHM). The UV-induced change in absorbance is calculated for each time delay $\Delta t$ as $\Delta A = -log_{10}(I_{\text{pump-on}}/I_{\text{pump-off}})$ where $I_{\text{pump-on}}$ is the XUV spectrum recorded in the presence of the pump beam, while $I_{\text{pump-off}}$ is recorded with the pump beam blocked using a chopping wheel. The time delay between each pump and probe pulse is varied between -1000 fs and +10,000 fs using a pair of mirrors mounted on a mechanical delay stage, with positive $\Delta t$ defined as the pump pulse preceding the probe pulse.

Spectral and temporal calibration of the setup is performed by filling the sample cell with xenon gas at pressures of ≪1 bar as previously described,[25] obtaining a FWHM of the instrument response function (IRF) of 110 ± 15 fs. The UV pump pulse duration is significantly longer than the XUV probe pulse duration.

**2.2. *Ab initio* calculations**

Density functional theory (DFT) calculations on 2-iodothiophene were performed with the Q-Chem 5 package,[26] using the cam-B3LYP[27] functional. Iodine N-edge (4d inner-shell) excitation energies and oscillator strengths were computed with a restricted open-shell Kohn-Sham (ROKS) approach,[28,29] using the aug-cc-pωCVTZ[30] basis on I (along with its associated pseudopotential) and the aug-cc-pVDZ[31,32] basis on other atoms. Time-dependent DFT (TDDFT) calculations on 2-iodothiophene were also performed with the def2-TZVPD basis[33] (utilizing the corresponding pseudopotential for I[34]), for computing valence excitations. Valence excitation energies obtained with ROKS (using the same basis set and pseudopotential as TDDFT) agreed very well with TDDFT, but slightly less so with the Tamm-Dancoff approximation (TDA).[35] All TDDFT calculations, therefore, were done without TDA, unless noted otherwise.

Fewest switches surface hopping[36] (FSSH) *ab initio* molecular dynamics calculations were also carried out, using TDDFT and the smaller def2-SV(P) basis [33]. The FSSH calculations started on the $\pi \rightarrow \pi^*$ electronic excited state, at the optimized (cam-B3LYP/def2-SV(P)) ground state geometry and the associated ground state zero-point energy. Further details about these calculations including calibration information are provided in the supplementary material (SM).

## 3. Results

The transient absorption spectra shown in Fig. 1A are obtained by integrating the measured $\Delta A(E, \Delta t)$ over separate pump-probe time-delay ranges as indicated in the figure legend. The pump pulse intensity in these measurements is approximately 3 TW/cm². Beyond ~50 eV, the spectra are dominated by negative changes in absorbance, which are due to depletion of the parent molecule concentration by UV pump excitation, probed by transitions from the iodine core $4d_{5/2}$ and $4d_{3/2}$ orbitals into molecular $\sigma^*$ and $6p$ orbitals. To distinguish core probe excitations from the valence excitations discussed to this point, we use the reverse notation. In particular, the $\sigma^* \leftarrow 4d_{5/2}$ transition is centered near 50.9 eV, $\sigma^* \leftarrow 4d_{3/2}$ near 52.6 eV, and $6p \leftarrow 4d$ transitions appear at $\gtrsim$55 eV. Positive contributions overlap with parent depletion in the Rydberg $6p$ window arising from atomic iodine $6p \leftarrow 4d$ transitions; one can be clearly distinguished near 55.5 eV. Below ~50 eV, UV-induced absorbance changes are positive, indicating the emergence of new species. Their spectral distribution evolves from essentially a single broad (~2 eV FWHM) feature at ~47.3 eV for the smallest pump-probe delays ($\Delta t$ = –30±20 fs, blue), to a series of narrow peaks at later times.

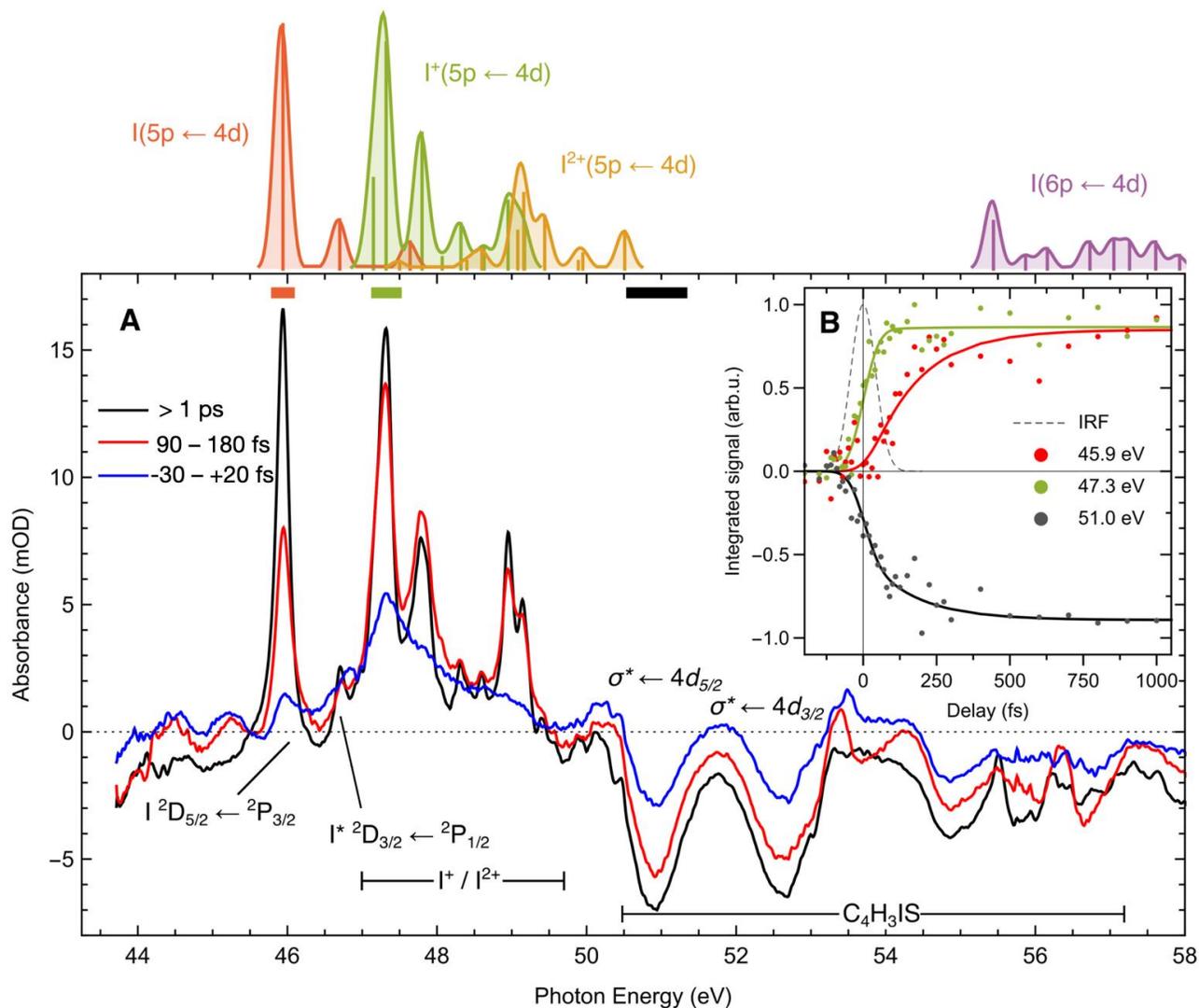

Figure 1. **(A)** UV pump – XUV probe transient absorption spectra at the iodine $4d$ edge of 2-iodothiophene. Negative-going signal at energies >50 eV corresponds to depletion of the parent molecule. At very early times (-30 – +20) fs (blue), a short-lived species is observed as a broad feature that peaks at 47.3 eV. At intermediate delays (90–180) fs (red) pronounced asymmetry in the neutral I atom peak near 46 eV is observed, while after >1 ps (black) the asymptotic products are formed. Color-coded bars near the top of panel (A) indicate the energies selected to illustrate time-dependent spectral trends in panel B. **(B)**

Circular points are measured data using the same color codes (energy specified in legend). The solid lines correspond to the result of a global fit analysis. The instrument response function (IRF) measured in xenon is overlaid (dashed gray). Literature $np \leftarrow 4d$ absorbance for both iodine atoms and ions are shown at the top of the figure, as a stick spectrum and after Gaussian convolution.[20]

To guide the assignment of the many sharp absorption features observed at long delays, literature $np \leftarrow 4d$ absorbance for both iodine atoms and ions is shown at the top of the figure, as stick spectra and after Gaussian convolution to match the experimental energy resolution (~100 meV).[37] Neutral I atoms are probed by $4d^9\ 5s^2\ 5p^6 \leftarrow 4d^{10}\ 5s^2\ 5p^5$ core-to-valence excitations, where both the core and valence orbital energies are split by spin-orbit coupling. Ground state I atoms are probed by the $^2D_{5/2} \leftarrow {}^2P_{3/2}$ and $^2D_{3/2} \leftarrow {}^2P_{3/2}$ transitions into spin-orbit-split core-excited states at 45.94 eV and 47.64 eV, respectively, while (valence) spin-orbit excited I* atoms are observed by the $^2D_{3/2} \leftarrow {}^2P_{1/2}$ transition at 46.70 eV.[37] The proportion of spin-orbit excited atoms $\Phi_{I^*} = [I^*]/([I^*] + [I])$ is estimated as 0.23 ± 0.03 (taking into account the different XUV oscillator strengths [19,37]), which indicates that formation of spin-orbit excited iodine fragments is a fairly minor channel. Multiple sharp features in the range 46.9–49.5 eV correspond to absorption by I$^+$, indicating contributions from multi-photon processes.[37] Contributions from I$^{2+}$ products appear to be minor at this pump intensity (see also Fig. S8 in the SM). The ionization energy of 2-iodothiophene is 8.5 eV.[38] As one-photon absorption is resonant with bound excited states and two UV pump photons provide 9.2 eV, the probability for driving Resonance Enhanced Multi-photon Ionization (REMPI) processes is significant. Throughout this work, we primarily consider the major neutral channel pathways, assuming they are independent of ionization channels, even though some depletion of neutral species might occur by ionization. A more detailed discussion of multi-photon contributions to the observed signals is provided in the SM.

Interestingly, before the appearance of the sharp atomic features, only approximately one third of the total parent molecule population depletion has occurred (blue trace). At intermediate delays, $\Delta t$ = 90–180 fs (red), the neutral I atom signal has reached approximately half the asymptotic intensity and a pronounced asymmetry in the lineshape is observed on the high-energy side of the peak. In contrast, I$^+$ ions form significantly faster than neutral I atoms and have already reached their asymptotic yield at this point in time.

The time-dependent behavior of the XUV transient absorption spectra is explored in more detail in Fig. 1B by inspecting key spectral regions marked by color-coded horizontal bars in the upper region of Fig. 1A. The circular data points in Fig. 1B are obtained by integrating the time-dependent signals across 0.1-0.5 eV wide spectral regions (indicated by the widths of the bars in Fig. 1A), and their respective maxima are normalized to unity. The same color codes are used in Fig. 1A and 1B. Solid curves are the result of a fit procedure described in the following section. The time-dependent depletion of the parent 2-iodothiophene population, as indicated by the I(4$d$) core-to-valence transition near 51 eV, is shown in black. A crucial observation is that following UV excitation, the 2-iodothiophene population bleach continues to increase up to delays of ~1 ps. Absorption changes in this energy window correspond to molecular, rather than atomic, iodine species and conclusively show that dissociation requires up to ~0.5-1 ps to complete. The emergence of neutral I atoms as indicated by the peak at 45.94 eV is shown in red. The appearance of neutral I atoms is clearly delayed compared to the IRF (gray, dashed), beginning to rise only after ~100 fs have elapsed and gradual changes in the asymptotic yield can be seen up until ~1 ps. Time delays >1 ps do not influence the atomic I peak intensity, position or lineshape, indicating that changes to the valence electronic environment have ceased at this point (see Fig. S1 in the SM). Signal integration around 47.3 eV illustrates the emergence of the most prominent I$^+$ transition and is shown in green. The appearance of I$^+$ is fast, reaching its maximum in less than 150 fs and is likely IRF limited.

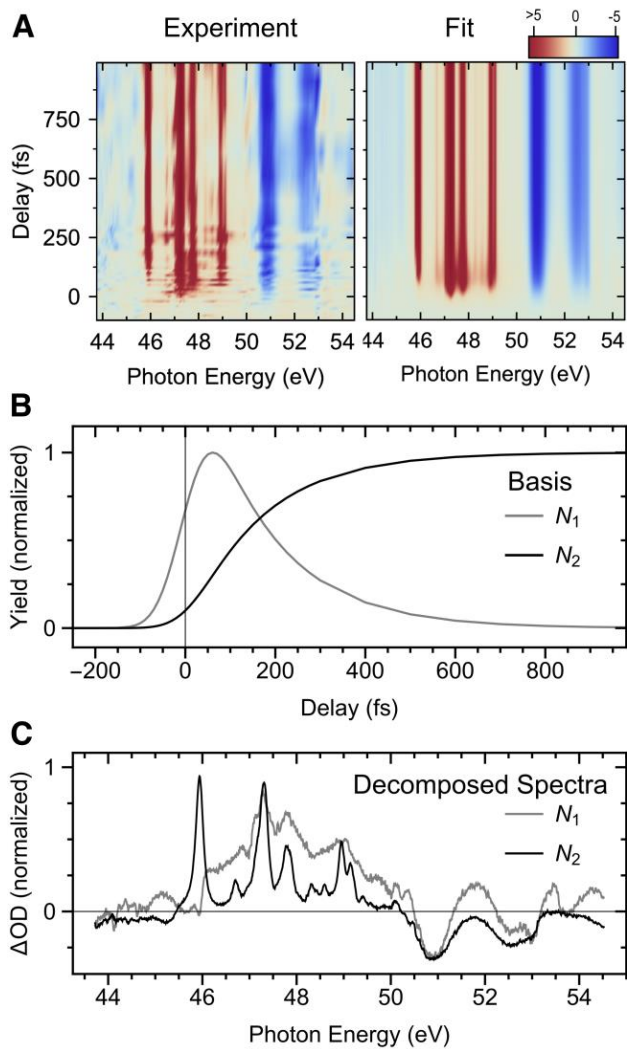

Figure 2. Global fit analysis of the time-resolved XUV absorption spectrum of 2-iodothiophene. **(A)** Measured (left) and global fit (right) data as a function of XUV absorption energy and pump-probe time delay. **(B)** The time basis set is defined by a two-state model where each state is sequentially populated. $N_1$ is populated by a Gaussian population function and depopulated into $N_2$ by a rate constant determined from the fit. **(C)** Decomposed spectra.

# 4. Analysis

## 4.1. Global fit procedure

A global fit procedure based on the standard least-squares algorithm is used to decompose overlapping spectral features of the time-resolved spectra into their transient spectral contributions. The left panel in Fig. 2A shows a false-color representation of the measured $\Delta A(E, \Delta t)$ trends as a function of energy (horizontal) and time (vertical) up to $\Delta t$ = 1000 fs for a pump pulse intensity of 3 TW/cm². Signal depletion appears in blue, enhancement in red as indicated by the color bar. The acquired transient absorbance signal may be regarded as a data matrix (TXA) of dimensionality ($t \times e$), where $t$ denotes the number of sampled time delays and $e$ the number of sampled energies. Assuming that the shape of the absorption spectrum of each transiently populated state is time-independent and the population of each state is time-dependent, the data matrix can be decomposed into a spectrum matrix, $E(n \times e)$, and a time matrix, $T(n \times t)$, with $n$ denoting the number of states. We caution that these states do not necessarily correspond to actual electronic states, but should be considered as dominant spectral components, similar to the concept underlying singular value decomposition (SVD). A variety of different kinetic models have been evaluated to describe the measured data with the goal of achieving a good description with the fewest number of free fit parameters. A two-state model with time-dependent populations $N_1$ and $N_2$ that are sequentially populated, is able to reproduce the observed dynamics well. Within this model, spectral component $N_1$ is initially populated by the pump pulse and decays into component $N_2$ with a rate $k_1$. The model can be described by the following system of differential equations:

$$\frac{dN_1}{dt} = g(t - t_0, \sigma) - k_1 N_1(t) \tag{1}$$

$$\frac{dN_2}{dt} = k_1 N_1(t) \tag{2}$$

Here, $g(t, \sigma)$ indicates a Gaussian-shaped initial population function accounting for the instrument response (of temporal width $\sigma$) and $k_1$ is the rate for transitions between states 1 and 2. The rate $k_1$ (and two time-independent amplitude factors) are free fit parameters, while $g(t,\sigma)$ is fixed to a Gaussian fit of the measured IRF. This approximation is feasible as the IRF is dominated by the pump pulse duration.

The result of the fit procedure is shown in the right panel of Fig. 2A. The time-dependent behavior of the fit and the experimental data are compared in detail in Fig. 1B. Note that the fit is applied to all experimental delays acquired ($\Delta t = -1000$ fs to 10,000 fs). The fit provides a good description of the measured $\Delta A(E, \Delta t)$ trends, including the different appearance times of different spectral features, despite having used only a single rate parameter in the global fit. With an IRF FWHM of 110 fs, a timescale of $\tau_1 = 1/k_1 = 165 \pm 30$ fs is found to capture all observed dynamics (also see Fig. S1). The temporal basis set, corresponding to the time-dependent population functions $N_1(t)$, $N_2(t)$, is shown in Fig. 2B. The corresponding decomposed spectral components are shown in Fig. 2C. The contributions of $N_1(t)$ and $N_2(t)$ to each of the I (red), I$^+$ (green) and parent 2-iodothiophene (black) features identified in Fig. 1B are decomposed in Fig. S2. A more complex model that treats neutral dissociation and ionization separately is discussed in the SM. From this model we can estimate the ratio between neutral dissociation and ionization to be approximately 1 at this pump pulse intensity. The timescale for neutral atom formation remains unchanged within the fit uncertainty, $\tau_{\text{neutral}} = 160 \pm 30$ fs.

### 4.3. Theoretical calculations

XUV transient absorption at the I(*4d*) edge probes the photoinduced chemistry from the viewpoint of the I atoms since only transitions from well-localized I*(4d)* inner-shell orbitals to valence orbitals exhibit notable oscillator strengths. However, interpreting the spectral trends presented above requires theoretical predictions of both the UV excitation induced coupled electronic nuclear dynamics, as well as the XUV spectral fingerprints of initially excited electronic configurations and potential candidates for molecular intermediates and final reaction products. Within a simplified picture, UV excitation lifts an electron from an occupied molecular orbital (MO) to an unoccupied MO, resulting in two singly occupied orbitals, SOMO1 and SOMO2.

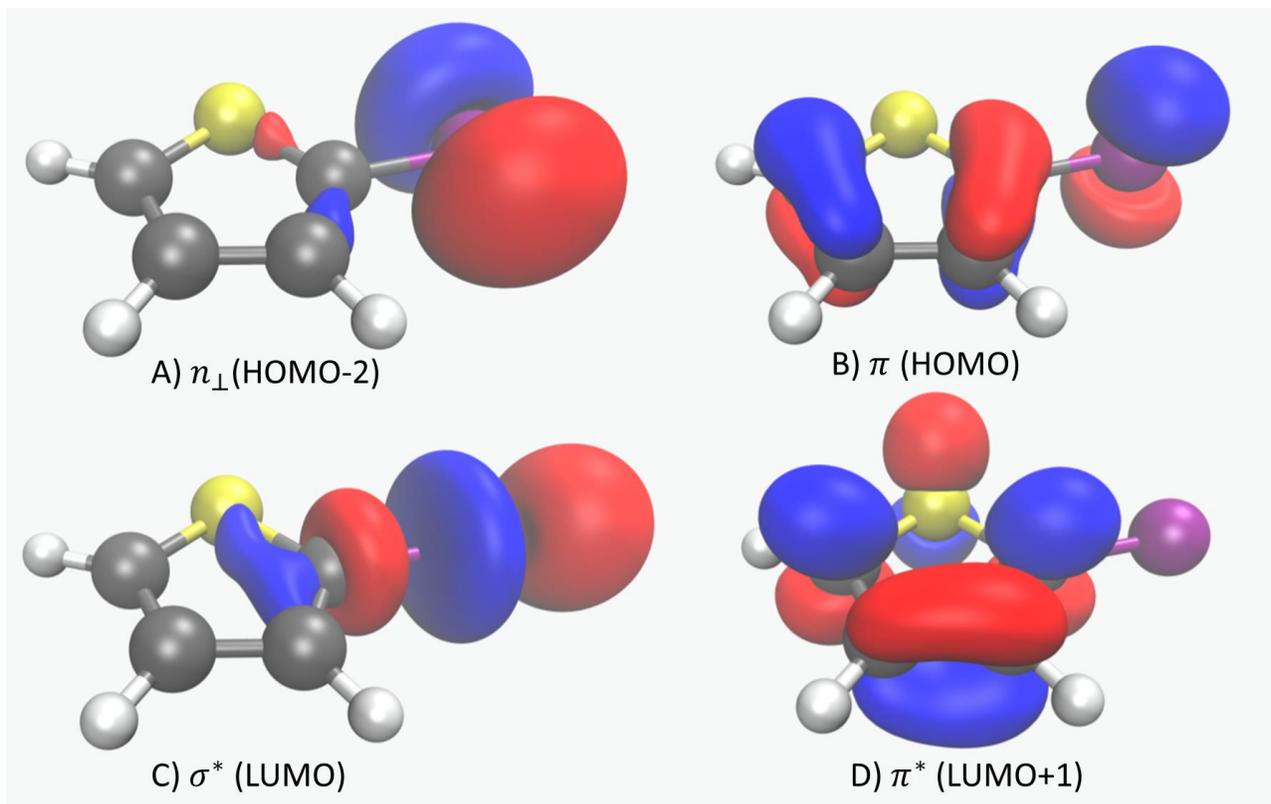

Figure 3. The relevant $S_0$ state orbitals of 2-iodothiophene at $S_0$ geometry (cam-B3LYP/def2-TZVPD, using an isovalue of 0.05 in VMD [39]). The HOMO-1 level is mostly an S 3p lone pair which is not involved with low lying excitations (<5.8 eV) and is thus not shown.

DFT calculations on ground state 2-iodothiophene indicate that the highest occupied MO (HOMO) is of π character, with contributions from both the thiophene aromatic system and the $n_\pi$ lone pair on the I atom (see Fig. 3B). On the other hand, the lowest unoccupied MO (LUMO) is a C-I σ* orbital (see Fig. 3C). TDDFT calculations predict that the lowest electronically excited state ($1^1A''$) is of π→σ* character, with an energy of 4.4 eV. This, however, does not have appreciable oscillator strength ($f < 10^{-4}$) from the ground state and is essentially dark. The lowest state with significant oscillator strength ($f > 0.1$) is of π→π* character ($3^1A'$), at 5.3 eV energy (which agrees reasonably well with the experimental band maximum at 239 nm/5.2 eV). In a simplified picture, this transition corresponds to excitation of an electron from the HOMO to the LUMO+1 (see Fig. 3). There is also the possibility of an $n_\perp \to \sigma^*$ excitation (i.e., an excitation from the I atom lone-pair spatially oriented perpendicular to both the C-I bond and the π system, see Fig. 3A) at 5.2 eV ($2^1A'$), which slightly mixes with the π→π* state, but this has significantly smaller oscillator strength (<1/6 of the $3^1A'$ state). We note that a closer inspection of the diabatic π→π* and $n_\perp \to \sigma^*$ contributions to the adiabatic $2^1A'$ and $3^1A'$ states indicates that the predominant oscillator strength in the adiabatic $2^1A'$ state is due to a partial π→π* contribution. Thus, in a diabatic picture, contributions from $n_\perp \to \sigma^*$ excitations are probably even smaller than suggested by the adiabatic state oscillator strengths displayed in Table 1. For convenience, we henceforth refer to these excited electronic states by their dominant diabatic character. Higher energy excitations (which are 5.8 eV or higher) are unlikely to be involved at the 268 nm (4.6 eV) pump energy and are not considered. We also note that ROKS gives virtually the same excitation energies as TDDFT for the π→σ*, $n_\perp \to \sigma^*$, and π→π* excitations. Furthermore, the TDDFT vertical excitation energies agree well with equation of motion coupled cluster singles and doubles, with perturbative triples (EOM-CCSD(fT)) calculations,[40] indicating the suitability of the cam-B3LYP functional (as discussed in the SM).

| Excited State (dominant character) | Energy relative to ground state (and TDDFT oscillator strength) | Lowest iodine N-edge 4d excitation energy (and ROKS oscillator strength) | % I contribution to SOMO1 |
|---|---|---|---|
| $1^1A''$ ($\pi \to \sigma^*$) | 4.4 eV (f ≈ $3 \times 10^{-5}$) | 46.4 eV (f=0.039) | 47 |
| $2^1A'$ ($n_\perp \to \sigma^*$) | 5.2 eV (f ≈ 0.026) | 45.6 eV (f=0.084) | 97 |
| $3^1A'$ ($\pi \to \pi^*$) | 5.3 eV (f ≈ 0.156) | 47.5 eV (f=0.018) | 27 |

Table 1: Characteristics of the three lowest valence singlet excited states. Percentage I contributions were obtained from SOMOs optimized via ROKS for the valence excited state, using subshell Mulliken analysis.[41] All calculations were done at the optimized $S_0$ ground state geometry.

ROKS calculations can be used to model iodine N-edge (4d inner-shell) XUV spectroscopy and determine whether this experimental technique can distinguish between the three lowest valence excited states at the Franck-Condon (unrelaxed ground state) geometry. ROKS accurately predicts the $\sigma^* \leftarrow 4d_{5/2}$ excitation energy of the $S_0$ state to be at 50.8 eV (vs 50.9 eV from experiment), indicating its suitability for modeling I(4d) spectroscopy (see also discussion in the SM). The lowest energy I(4d) excitations for the valence excited states are to the lower energy singly occupied molecular orbital (SOMO1), which was formerly doubly occupied (i.e., excitation into the vacancy created by the valence electronic excitation). The computed SOMO1←I(4d) excitation energies for the three states of interest are provided in Table 1. The lowest N-edge excitation energy for the $n_\perp \to \sigma^*$ state is, unsurprisingly, close to the atomic I(5p)←I(4d) excitation (which is computed to be at 45.5 eV vs 45.9 eV from experiment, see SM). The SOMO1 levels for both the $\pi \to \sigma^*$ and $\pi \to \pi^*$ excitations are derived from the ground state HOMO, but the lowest 4d inner-shell excitation energy for the former is lower by an eV. This is largely a consequence of differing contributions from the I atom to the SOMO1, as orbital relaxation via ROKS indicates that the $\pi \to \sigma^*$ SOMO1 has substantially greater I atom contribution than the $\pi \to \pi^*$ SOMO1 (which is more localized on the ring). The π→π* SOMO1 is consequently less stabilized by the core hole localized entirely on I, leading to greater excitation energy. This interaction between the excited electron orbital and the core hole is sometimes described in terms of a core exciton binding energy, as discussed in more detail in a previous inner-shell transient absorption study.[25] The XUV excitation oscillator strengths also decrease with decreasing I atom contribution to the SOMO1. We note that the computed lowest I 4d inner-shell excitation energies for π→σ* and $n_\perp$→σ* states are very close to the difference between the lowest I 4d inner-shell excitation energy of $S_0$ 2-iodothiophene (σ*←4d, 50.9 eV) and the valence excitation energies of the two states (4.4 eV/5.2 eV, respectively), which is expected, as the energy of an electronic state is a state function and the molecular geometry remains the same. A similar argument could be made for the π→π* state based on the π*←4d excitation of $S_0$ 2-iodothiophene (lowest multiplet computed at 52.8 eV), but this state is not evident in the experimental static spectrum as the excitation has rather low oscillator strength due to the π* orbital having virtually no contribution from I (Fig. 3D) and overlaps with the higher energy multiplet of the σ*←4d process.

The next higher energy excitations for these species (aside from the spin-orbit multiplet) are ~50 eV (to higher energy σ*/ π* levels) and overlap with the ground state absorption of 2-iodothiophene. The 45-50 eV region of the transient XUV absorption spectrum is therefore well suited to indicate which of the lowest three excited states are forming upon 268 nm excitation of 2-iodothiophene.

## 5. Discussion

The experimental transient absorption spectrum at pump-probe delays <20 fs has a prominent feature at 47.3 eV (see blue trace in Fig. 1A). The results of the *ab initio* calculations presented in Table 1 guide assignment of this feature to absorption from a π→π* excitation as the dominant initial electronic state. This finding is in good agreement with the much larger oscillator strength for the ground state to valence π → π* excitation compared to the other two possible 268 nm UV induced transitions. There are some small features at lower energies that might correspond to a small fraction of the molecules being excited to the π→σ* and/or $n_\perp$→σ* states, but not in appreciable amounts, especially considering that these states have larger I 4d inner-shell oscillator strengths. We compute the lowest I 4d inner-shell absorption feature of the 2-iodothiophene cation ground electronic state to be located at 48.0 eV (SOMO←4d) for the unrelaxed $S_0$ geometry and at 48.3 eV for the relaxed cation geometry. The main 47.3 eV signal thus cannot arise from the cation ground state, although some portion of the broad tail at higher energies (such as the shoulder at ~48.5 eV) could be indicative of the presence of the cation.

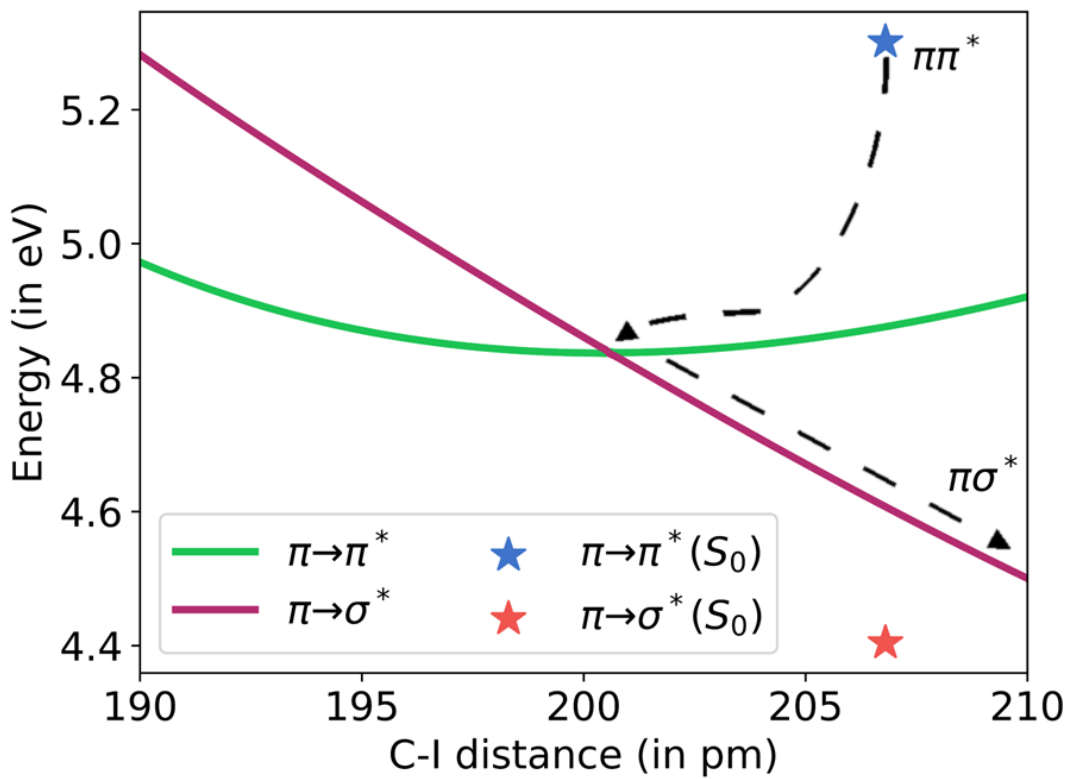

Figure 4. Variation in the (TDDFT) energies of the π→π* and π→σ* states vs C-I bond distance (the 2-thienyl unit being kept constrained to the MECP geometry). The energies of the two states at the $S_0$ geometry are also provided (blue star for π→π* and orange star for π→σ*) for context. All energies are relative to the $S_0$ state at equilibrium geometry.

Our theoretical calculations show that an initial π→π* excitation is consistent with the experimentally observed, relatively slow appearance (160 fs) of I atom signals, as typical C-I bond direct dissociation 1/e risetimes are 25-50 fs. [20,21] An initial excitation to the C-I σ* orbital is expected to lead to immediate cleavage of the C-I bond,[7,42] typically producing appearance times of neutral iodine fragment signals on sub-100 fs timescales.[19,20] On the other hand, a π→π* excitation can require significant electronic and nuclear relaxation before the I atom can depart. We note that molecular rotation and associated changes in alignment-dependent transition moment oscillator strengths are not a significant contributor to the trends observed in this work. Based on the smallest eigenvalue of 2-iodothiophene's moment of inertia tensor (918768 amu pm$^2$ from our DFT optimized $S_0$ geometry, corresponding to a rotation axis essentially coincident with the C-I bond) and a sample at room temperature, we estimate the shortest period of rotation to be ~3.8 ps, which is more than an order of magnitude longer than the characteristic I atom appearance time in the experiment. TDDFT optimization of the geometry of the π→π* excited state (starting from the $S_0$ geometry) reveals a conical intersection (CI) between the π→π* and π→σ* states at a reduced C-I bond length compared to the neutral molecule ground state (see Fig. 4). This behavior can be understood in terms of the character of the orbitals involved and their dependence on the C-I bond length. The HOMO π level has an antibonding interaction between the I 5p orbital and the ring π system, as can be seen from Fig. 3B. The π→π* electronic excitation reduces this interaction and thereby contraction of the C-I bond lowers the excited state energy. However, contraction of the C-I bond length leads to an increase in the energy of the σ* level, causing the energy of the π→σ* state to increase. The gap between the π→π* and π→σ* states, therefore, shrinks with C-I bond contraction, as depicted in Fig. 4. The TDDFT optimized minimum energy crossing point (MECP) has a C-I distance of 201 pm (vs 207 pm in the DFT optimized ground state), with the energy of the π→π* state decreasing by 0.47 eV and the energy of π→σ* state increasing by 0.44 eV relative to the ground state geometry, as shown in Fig. 4. While the precise computed geometry of the calculated MECP can be somewhat theory dependent, the gap between the π→π* and π→σ* states decreases significantly as the nuclei relax on the π→π* surface. This indicates that an initial π→π* excitation can cross over to the π→σ* state after a time delay for nuclear relaxation, followed by cleavage of the C-I bond on the repulsive π→σ* surface. Notably, this channel precludes any need for C-S bond cleavage, allowing the thienyl ring to stay intact over the course of the reaction. However, it is worth noting that the MECP geometry has an elongated C-S distance relative to the ground state (178 pm vs 171 pm) due to the antibonding interaction between S and the ring C atoms in the π* level. Ring opening thus remains a possible pathway for excited 2-iodothiophene, even if it is a minor contributor. We note that, for UV excitation of 2-iodothiophene in the energy region probed here, Marchetti et al. proposed the existence of similar pathways, i.e., an initial π→π* excitation with bonding character, followed by coupling to dissociative *n*→σ*, π→σ* states in both the C-I (dissociation) and C-S (ring opening) coordinates.[7] These insights were based on a combination of fragment velocity map imaging experiments and *ab initio* electronic structure calculations. The combined experimental-theory results presented here support this general picture and provide additional insight through time-resolved monitoring of both the character of the initially excited state as well as the subsequent temporal evolution of the molecule toward the emergence of atomic fragments by ultrafast inner-shell spectroscopy.

The actual relaxation channels available to the π→ π* excitation are investigated by means of *ab initio* molecular dynamics (AIMD) calculations (starting from the ground state geometry) utilizing TDDFT and Tully's FSSH approach.[36] Out of 100 trajectories, 67 led to C-I cleavage without thiophene ring opening, indicating that this is indeed the major channel. The ring opened via C-S bond breaking in the other 33 cases, albeit without departure of the I atom. Considering the uncertainty associated with the finite number of trajectories sampled, the AIMD results indicate C-I dissociation in 67±5% of the cases, to one standard deviation. We note that Marchetti et al. proposed the existence of low energy forms with both C-I and C-S bonds breaking. However, such products require the C-S bond to be broken first, followed by H migration that releases sufficient energy to also permit breaking the C-I bond. In this context, it is worth noting that no strong signals for the initial ring opened product (without further structural reorganization such as H migration) were observed in the experimental iodine I 4d inner-shell spectrum, with the lowest absorption energies of this biradical form computed to be at 49.3 eV (f ~ 0.0024) and 49.8 eV (f ~0.020), corresponding to excitations to the singly occupied biradical orbitals. The lowest energy form reported by Marchetti et al. corresponds to an H migration in this ring opened product from the C atom in the 5- position to the C in the 2- position. The resulting low-energy closed-shell molecule however has a computed I 4d inner-shell absorption profile very similar to the parent 2-iodothiophene. Formation of this structure in appreciable quantities therefore would correspond to a reduction of ground state bleach that is not observed in practice. We further note that our AIMD trajectories do not indicate the onset of any H migration prior to the first bond breaking. It is nonetheless worth emphasizing that the I atom is not necessarily the preferred local reporter of dynamics on the thienyl ring, which would be more clearly revealed by inner-shell studies utilizing the S(2p) and/or C(1s) electrons. Efforts are underway to extend the reach of the experimental setup to excitations involving these much more strongly bound electronic orbitals.

It is natural to wonder at what point the I 4d inner-shell absorption signal for the π→σ* state switches from "molecular" to "atomic" in character over the course of the dissociation process. There are three occupied orbitals that undergo substantial change of character over the bond cleavage. The first is the C-I σ bonding orbital, which changes from a shared MO in the bonding region to a localized 5p lone pair on the I atom at the dissociation limit. This cannot be detected by XUV absorption, on account of this level being doubly occupied throughout the dissociation process. On the other hand, the π SOMO (SOMO1) loses all ring character and becomes the singly occupied 5p orbital on atomic I, as shown on the left of Fig. 5. Additionally, the singly occupied σ* (SOMO2) loses all I character and becomes a sp$^2$ radical orbital on the C, as depicted on the right of Fig. 5.

| C-I separation (in pm) | $E_{ex}$ (SOMO1), in eV | $f$ (SOMO1) | % I contribution (SOMO1) | $E_{ex}$ (SOMO2), in eV | $f$ (SOMO2) | % I contribution (SOMO2) |
|---|---|---|---|---|---|---|
| S$_0$   207 | 46.4 | 0.039 | 47 | 50.6 | 0.027 | 66 |
| 220 | 46.4 | 0.033 | 38 | 49.5 | 0.033 | 59 |
| 230 | 46.2 | 0.031 | 37 | 48.9 | 0.034 | 56 |
| 235 | 46.1 | 0.045 | 52 | 48.6 | 0.024 | 50 |
| 240 | 46.0 | 0.049 | 57 | 48.3 | 0.021 | 45 |
| 250 | 45.8 | 0.056 | 65 | 47.8 | 0.015 | 38 |
| 260 | 45.7 | 0.064 | 74 | 47.5 | 0.009 | 30 |
| 265 | 45.7 | 0.073 | 83 | 47.3 | 0.006 | 24 |
| 270 | 45.6 | 0.076 | 88 | 47.2 | 0.004 | 20 |
| 275 | 45.6 | 0.080 | 91 | 47.1 | 0.002 | 17 |
| 280 | 45.6 | 0.086 | 93 | 47.0 | 0.002 | 15 |
| Dissociation | 45.5 | 0.086 | 100 | 55.3 | 0.000 | 0 |

Table 2: Excitation energies ($E_{ex}$) and oscillator strengths (f) for the lowest multiplet iodine 4d inner-shell excitations in the lowest π→σ* excited state of 2-iodothiophene. The percentage of the relevant SOMOs that arise from I atomic orbitals (estimated via Mulliken subshell analysis) is also reported. The geometries were optimized with TDDFT for each C-I separation, except for the S$_0$ geometry, which was used as is.

The evolution of the SOMO1 and SOMO2 orbitals from the parent molecule to the molecular radical + atomic fragment limits are accompanied by changes in the iodine 4d inner-shell absorption, as illustrated in Table 2. It lists the energies and oscillator strengths of I(4d) to SOMO1/SOMO2 excitations for the π→σ* excited state across various C-I distances, starting from the S$_0$ equilibrium bond length. The excitation energy for SOMO1 undergoes a redshift from 46.4 eV to 45.5 eV as the C-I distance increases, corresponding to an increase in the I atom contribution to SOMO1 (see Fig. 5, left, and percentage I contributions in Table 2) that leads to greater stabilization of the I (4d) excited state due to more negative core exciton binding energy. We tentatively assign the small high-energy shoulder of the atomic I($^2D_{5/2}$ ← $^2P_{3/2}$) peak at intermediate times (Fig. 1A) as a signature of this transition during the dissociation process. Perhaps more revealing, however, is the change in oscillator strength, which undergoes a significant increase between 230-280 pm C-I distance, essentially reaching the atomic I value within an extension of ~50 pm. Such an outcome is suspected to be the cause of amplitude changes in the well-isolated I($^2D_{5/2}$ ← $^2P_{3/2}$) peak at 45.9 eV, whose intensity roughly increases by double between intermediate (90-180 fs) and long (>1 ps) delays, as seen in Fig. 1A.

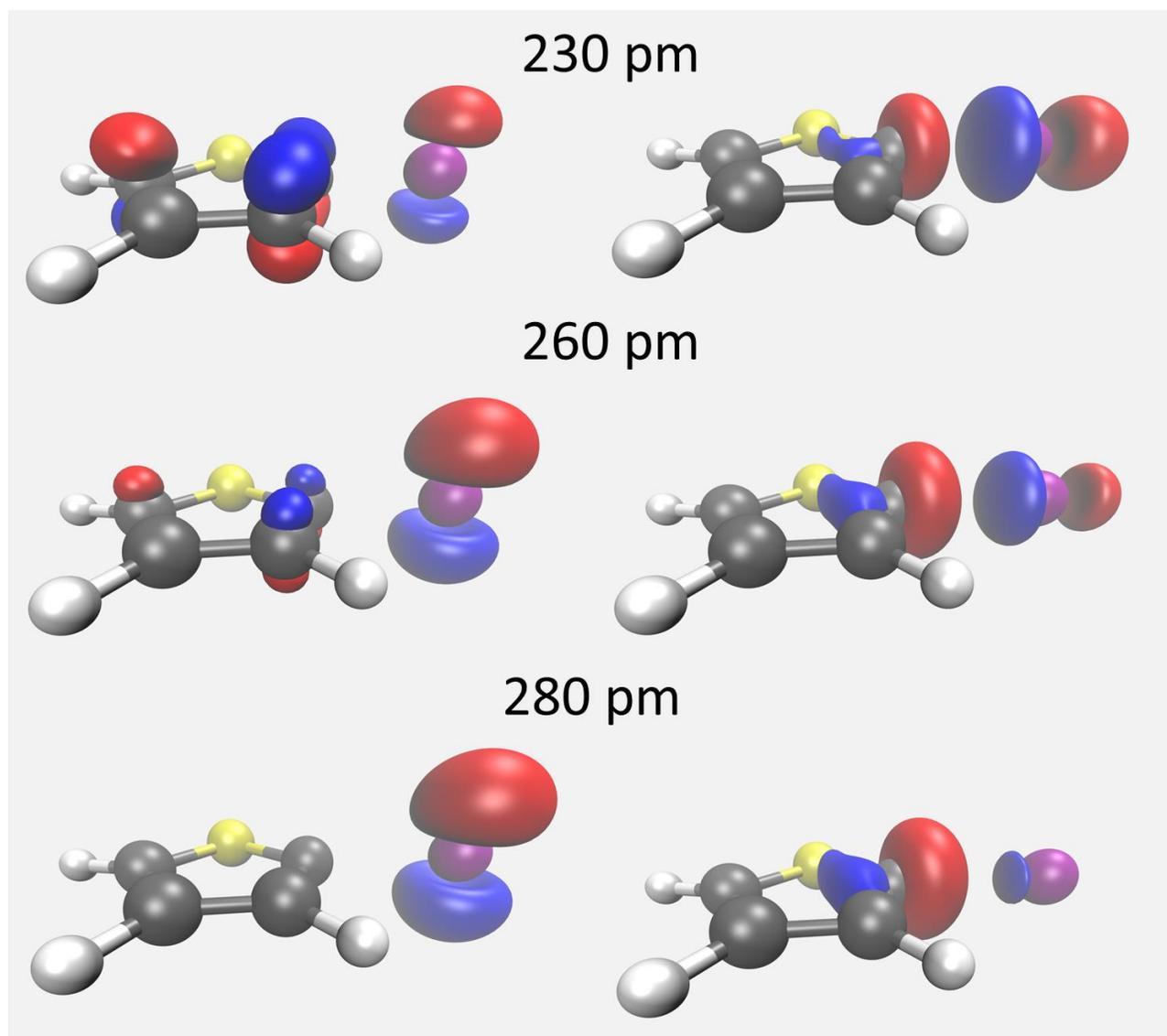

Figure 5. ROKS SOMO1 (left) and SOMO2 (right) for C-I separations of 240 pm (top), 260 pm (middle) and 280 pm (bottom), using isovalue=0.08 in VMD.[39] Localization on individual fragments is nearly complete between 230-280 pm.

The energies of the SOMO2←I(4d) excitations undergo more pronounced and complex variations than those of the SOMO1 excitation. They exhibit an initial lowering by ~4 eV upon stretching the C-I bond to 280 pm, but then increase by ~8 eV to the dissociation limit. This behavior can be understood by distinguishing between the molecular and fragment regimes along the bond dissociation pathway (see Fig. 5, right). Initially, SOMO2 is a σ* level delocalized over both C and I, with the energy of the orbital decreasing with increasing C-I distance. SOMO2←I(4d) excitation energies should thus decrease in this regime. However, the relative I contribution to SOMO2 starts to decrease at some point, going to zero at the dissociation limit. At this point, the SOMO2←I(4d) excitation is a charge-transfer from the I(4d) core to a vinylic $sp^2$ C radical orbital on the thienyl ring. This type of excitation has a high energy, computed to be 55.3 eV, but has zero oscillator strength. The final stage of the transition from the molecular to fragment regime for the σ* orbital should thus be marked by an increase in the core-level excitation energy, as an orbital localized on the C atom and populated by a charge-transfer excitation from the iodine atom would be less stabilized by electron-hole interaction than excitations more localized on the iodine atom. We note that this predicted increase in the SOMO2←I(4d) excitation energy at long separations is however not computationally observed at stretches up to 280 pm (as shown by Table 2), likely due to the relatively slow rate with which charge-transfer states increase in energy with increasing separation r (asymptotically, $O(r^{-1})$). More generally, the overlap of molecular excited state and ground state absorption energies leads to the relatively slow evolution of the black depletion curve in Figure 1B, which extends well beyond the instrument response as a result of intermediate molecular state signals partly compensating the depletion of the ground state spectrum.

The SOMO2←I(4d) oscillator strength should also decrease to zero with increasing C-I separation, as the SOMO2 level localizes on the thienyl ring. This behavior of the oscillator strength is computationally observed in Table 2, with only 7% of the initial oscillator strength persisting by 275 pm C-I stretch. Quantitatively, Mulliken subshell population analysis on ROKS solutions indicates that the percentage I contribution to the SOMO2 σ* level is 56% at 230 pm, 30% at 260 pm and 15% at 280 pm, indicating that C-I cleavage is essentially complete by 280 pm. A visual depiction of this transition can be seen in Fig. 5, right. A quantitative comparison of the computed trends for the SOMO2←I(4d) excitations with the experimental data is challenging due to the rapid decrease in oscillator strength corresponding to this excitation and overlap with the experimental signal for I+ in the 47-49 eV regime. However, the continuous, monotonic decrease of absorbance observed for energies around ~50 eV up to ~1 ps time delay is consistent with the loss of XUV absorption oscillator strength in this region as expected from the trends for I(4d) inner-shell transitions to the SOMO2 orbital shown in Table 2. Together with the findings for the SOMO1←I(4d) transitions described above, a consistent picture emerges, whereby the observed delayed rise of I fragments and decay of parent molecule signals, as illustrated in Fig. 1B, are signatures of the transition from the intact parent molecule to the thienyl radical / iodine atom pair limit, as the dissociation proceeds on the dissociative πσ* surface.

It is worth noting that the Coulson-Fischer point[43] for C-I stretches on the ground state surface is 290 pm, which should be an acceptable indicator of localization of bonded electron density (in the ground state) with a reasonable functional like cam-B3LYP.[44] The 230-280 pm range for the π→σ* thus tracks well with ground state behavior based on the above analysis. The aforementioned AIMD calculations indicate that an average time of 75-84 fs is needed to attain C-I bond dissociation. Specifically, the average time to stretch to 260 pm is 75 fs and 84 fs for 280 pm, over the 67 trajectories where the cleavage is seen. This appears to be somewhat smaller than the experimentally observed delay of ~160 fs (with both of these quantities corresponding to inverse rate constants associated with exponential decay). The difference likely stems from the use of a single nuclear configuration ($S_0$ geometry) for the AIMD calculations instead of the full nuclear wavepacket. Additional causes for the discrepancy can be limitations of the TDDFT calculations (including but not limited to challenges around the Coulson-Fischer point[45] and the small basis set used for the dynamics) and/or FSSH.

We have also investigated two alternative channels for I atom loss. The first is from a vibrationally excited (i.e., "hot") 2-iodothiophene molecule with 4.6 eV excess energy, corresponding to the pump photon energy. RRKM calculations indicated that the I atom loss rate by this mechanism would be 0.3 ns$^{-1}$, which is far too slow to be a significant channel. The other channel considered is loss from the "hot" open-ring product. This was estimated via AIMD calculations on the lowest triplet surface, starting from the equilibrium geometry of ring-opened 2-iodothiophene with 1.5 eV excess energy, corresponding to the energy difference between this structure and ground state 2-iodothiophene pumped by 4.6 eV extra energy. Out of 100 trajectories (that were run for 1.8 ps), 16 attained C-I distances of 260 pm, taking an average time of 1.3 ps. Each of these 16 trajectories also involved C-S bond formation to restore the ring, and no indication of H migration was observed. It thus appears that loss of I atoms from hot ring-opened 2-iodothiophene would require a few ps of time and is thus fairly slow as well. The singlet open-ring structure is likely to be even slower in cleaving the C-I bond, having less excess nuclear energy to begin with, on account of the singlet-triplet electronic gap. It is worth keeping in mind that these specific AIMD calculations were done on a single surface, and state crossings, especially to repulsive states, could affect rates. Nonetheless, it does appear that adiabatic I loss from vibrationally excited non-repulsive states is considerably slower than the experimentally observed timescales.

The results presented here for the 268 nm dissociation of 2-iodothiophene add interesting nuances to the rich literature on UV-induced photochemistry of haloalkanes and aryl halides. UV absorption in the A-band of iodoalkanes corresponds to n→σ* excitations from a nonbonding iodine *p* orbital to an antibonding σ*(C-I) orbital.[12] The dissociative nature of the n→σ* states leads to a broad, structureless absorption spectrum. For aromatic species, however, such as iodobenzene and 2-iodothiophene, smooth (n/π) →σ* spectra are superimposed by sharper features due to contributions from additional π→π* excitations.[7,46-48] The energetically overlapping forms of excitation have distinctly different consequences. While excitation into the antibonding σ*(C-I) orbital leads to very rapid, direct dissociation producing fast photoproducts and highly anisotropic fragment ejection patterns, π→π* excitations result in more complex predissociation pathways that can proceed on timescales comparable to molecular rotation.[12,14,23,49] Thus, a deeper understanding of the competition between the different excitation/relaxation pathways is warranted in order to gain a more complete picture of the UV photochemistry of halocarbon molecules.

A comparison of findings on the 266 nm induced dissociation dynamics in chemically related halocarbon compounds illustrates the complex nature of the problem. In $CH_3I$, 266 nm absorption corresponds to an n→σ* excitation parallel to the C-I bond, which is followed by very rapid (<100 fs) dissociation along the C-I coordinate.[19,20,22] At the same excitation wavelength, 2-iodothiophene and iodobenzene exhibit signatures of both (n/π) →σ* as well as π→π* excitations. The translational energy release, determined from ion images of ground spin-orbit state I atom products, is qualitatively similar for both molecules. Two different types of product channels form; one is translationally fast, and commonly associated with (n/π) →σ* excitation, and one translationally slow, typically attributed to an initial π→π* excitation followed by predissociation and deposition of substantial amounts of internal energy into the aryl radical. At λ = 266 nm, the fast channel is dominant, accounting for ~80 % of all I atoms for both 2-iodothiophene and iodobenzene.[7,15,18]

We note, however, that there is not necessarily a 1-to-1 correlation between the translational energy release of the I fragment and the initial excitation. For 266 nm excitation of iodobenzene, Sage and co-workers observed a distinct local minimum in the ion ejection anisotropy parameter of $\beta \approx 0.7$ at the peak of the fast channel, and an average value of $\beta \approx 1.1$ across the entire feature, both of which are substantially smaller than the theoretical limit of $\beta = 2$ for a parallel transition, and both are smaller than anisotropy parameter values for the same feature at other excitation wavelengths between ~220 nm and ~320 nm.[18] The authors speculated that π→π* excitations overlapping with the predominant π→σ* excitation lead to this reduced anisotropy either through opening perpendicular excitation channels and/or by slowing down (pre)dissociation dynamics, which would lead to smaller experimental $\beta$ values by rotational averaging. In a later femtosecond time-resolved I(4d) inner-shell transient absorption study of iodobenzene using 265 nm pump pulses, Drescher et al. observed the emergence of iodine ground state fragments within ~40 fs and a faster, apparatus limited decay of the parent molecule signature. In this case, the (slightly) delayed appearance of the iodine signal was interpreted as the result of an electronic reorganization required to transfer hole density after the predominant π→σ* excitation from the phenyl ring to the iodine atom, opening the I(5p) ← I(4d) XUV absorption channel that is used to probe the I fragment.

The combined experiment-theory results presented here for 2-iodothiophene indicate that a very different scenario is at play in this molecule after 268 nm absorption. The π→σ* excitation has negligible oscillator strength from the ground state while the π→π* configuration is clearly identified as the predominant excitation channel, compared to $n_\perp$ →σ*. This conclusion is supported through both the relative oscillator strengths of the different transitions as well as the excellent agreement of the predicted energy of the SOMO1 ← I($4d_{5/2}$) inner-shell transition in the π→π* configuration (Table 1) with the experimentally observed spectrum at the smallest time delays (Fig. 1A). The subsequent relaxation dynamics are marked by significantly longer timescales (~160 fs) compared to $CH_3I$ and iodobenzene (<100 fs) and they are clearly observed in both the parent molecular depletion as well as the fragment ion signals. The findings are consistently described by a predissociation pathway coupling π→π* to π→σ*, as shown in Fig. 4 and provide a direct window into the transition from the molecular to the atomic regime as illustrated in Fig. 5 and Table 2. In particular, the relatively slow emergence of the atomic I($^2D_{5/2}$←$^2P_{3/2}$) signal intensities and the slight energy shifts/shoulder in this signal at early times, as predicted in Table 2 for the SOMO1 ← I(4d) transitions, as well as the continued decay of absorption in the region of the undissociated molecular SOMO2 ← I(4d) transitions (~50 eV, Table 2), are evidence for a relatively narrow region of ~230-280 pm along the C-I coordinate where the transition from the united molecular to the separate fragment regimes is taking place. We emphasize that this insight critically depends on the combination of both femtosecond time-resolved inner-shell absorption data as well as *ab initio* theoretical predictions of core and valence transition energies and oscillator strengths.

The important role of the lowest energy π→π* excitation in the UV photochemistry of 2-iodothiophene has been pointed out before. Zhu and co-workers observed activation of several ring stretching modes in resonant Raman spectra upon excitation with ~240 nm - 253 nm radiation and, based on TDDFT calculations, assigned the A-band absorption in 2-iodothiophene, in particular near the maximum at ~243 nm, predominantly to π→π* excitations that corresponding to a HOMO→LUMO+1 promotion.[12] A combined experiment-theory study by Marchetti et al. using ion imaging across a large wavelength range of ~220 nm - 304 nm suggests that (n/π) →σ* excitations dominate at longer wavelengths, while π→π* excitations gain increasing importance toward shorter wavelengths.[7] The results presented herein strongly suggest that it is indeed the latter channel that dominates the UV-induced dissociation dynamics in 2-iodothiophene following 268 nm excitation. We note that ring opening, as discussed by Marchetti et al.[7] and observed in the structurally similar 2(5H)-thiophenone ($C_4H_4OS$) molecule by Pathak et al.,[3] is not detected in this study. A more detailed discussion of this aspect of the UV-induced dynamics is given in the SM.

## Conclusions

A combined experiment-theory study of UV-induced dissociation dynamics in 2-iodothiophene at an excitation wavelength of 268 nm has been performed. Femtosecond time-resolved I(4d) inner-shell to valence XUV transient absorption spectroscopy reveals the emergence of free iodine fragments on a timescale of 160 ± 30 fs, which is significantly slower compared to previously studied iodine loss in UV-excited iodobenzene and various alkyl iodides. In further contrast to previous ultrafast XUV transient absorption studies of these systems, the parent molecule signal of 2-iodothiophene is not depleted within the experimental time resolution (~110 fs FWHM), but undergoes changes on very similar timescales as the emerging iodine fragment signal extending out to ~1 ps. Combining the experimental findings with the results of DFT, TDDFT, and ROKS calculations of valence and inner-shell transition energies and oscillator strengths suggest the following scenario.

Based on valence transition oscillator strengths and the XUV absorption spectrum at the earliest UV-pump – XUV-probe delays, the initial excitation is predominantly of π→π* character, possibly with minor contributions from $n_\perp$→σ*, while direct π→σ* excitation is very unlikely. The more strongly bound character of the π→π* excited state(s) along the C-I coordinate compared to the $S_0$ ground state leads to an initial shortening of this bond length by 6 pm, where the molecular wavepacket reaches a conical intersection of the π→π* state with a dissociative π→σ* potential energy surface. Nonadiabatic transition to this repulsive state is followed by dissociation to produce a thienyl radical and an iodine atom. Notably, the transition between the intact parent molecule and the separated radical–atom pair regimes is particularly pronounced within a relatively short range of C-I distances between 230 and 280 pm. Within this range, the cross section of the SOMO1 ← I(4d) transition increases significantly due to rapid localization of the SOMO1 on the iodine atom, where it emerges as a (singly occupied) I(5p) orbital. In other words, switching between the molecular and the atomic regimes occurs ~160 fs after UV excitation and at a C-I bond length of ~ 230-280 pm. This conclusion is corroborated by the theoretical prediction and experimental observation that the molecular absorption signal in the energy range associated with the SOMO2 ← I(4d) transition exhibits a depletion on the same timescale as the emergence of the atomic fragment. The atomic site-specificity of the transient inner-shell absorption signals is a key enabling factor for this remarkably detailed picture of the coupled electron-nuclear dynamics during the dissociation process. Alternative dissociation channels, such as hot ring-open product and hot ground state dissociation appear too slow (~ps - ns timescales) compared to the experimental observations to play any significant role in C-I bond fission.

Ring-opening, as suggested by related studies on 2-iodothiophene and 2(5H)-thiophenone, is not experimentally observed. Nevertheless, future studies using higher pump photon energies and probing in the range of the sulfur 2p edge or carbon 1s edge are warranted to gain better insight into this energetically allowed relaxation channel.

## Supplementary Material

See supplementary material for additional details of the global fit model and quantum chemistry calculations.

## Author Information

### Corresponding Author

*E-mail: ogessner@lbl.gov

## Author Declarations

### Author Contributions

**Benjamin W. Toulson:** Conceptualization (equal), Data curation (lead), Formal analysis (lead), Investigation (experiment, lead), Methodology (equal), Validation (lead), Visualization (lead), Writing - original draft (lead), Writing – review and editing (equal).

**Diptarka Hait:** Data curation (lead), Formal analysis (lead), Investigation (theory, lead), Methodology (equal), Software (lead), Visualization (lead), Writing - original draft (lead), Writing – review and editing (lead).

**Davide Faccialà:** Investigation (experiment, supporting), Writing - original draft (supporting), Writing - review and editing (supporting).

**Daniel M. Neumark:** Funding acquisition (supporting), Writing – review and editing (supporting).

**Stephen R. Leone:** Funding acquisition (supporting), Writing – review and editing (equal).

**Martin Head-Gordon:** Funding acquisition (equal), Methodology (equal), Resources (equal), Software (supporting), Supervision (theory, lead), Writing – review and editing (equal).

**Oliver Gessner:** Conceptualization (equal), Funding acquisition (equal), Methodology (equal), Validation (supporting), Project Administration (lead), Resources (lead), Supervision (experiment, overall, lead), Writing- original draft (supporting), Writing – review and editing (lead).

**Conflict of Interest**

M.H.-G. is a part-owner of Q-Chem, which is the software platform utilized for quantum chemical calculations in the present work.

# Acknowledgements

This work was supported by the Atomic, Molecular, and Optical Sciences program of the U.S. Department of Energy, Office of Basic Energy Sciences, Chemical Sciences, Geosciences and Biosciences Division, through Contract No. DE-AC02-05CH11231.

# Data Availability Statement

The data that support the findings of this study are available from the corresponding author upon reasonable request.

# Probing C–I Bond Fission in the UV Photochemistry of 2-Iodothiophene with Core-to-Valence Transient Absorption Spectroscopy

## Supplementary Material

Benjamin W. Toulson,[1,*] Diptarka Hait,[1,2,a,*] Davide Faccialà,[3] Daniel M. Neumark,[1,2] Stephen R. Leone,[1,2,4] Martin Head-Gordon,[1,2] and Oliver Gessner[1,⊥]

[1]Chemical Sciences Division, Lawrence Berkeley National Laboratory, Berkeley, California 94720, USA

[2]Department of Chemistry, University of California, Berkeley, California 94720, USA

[3]CNR-Istituto di Fotonica e Nanotecnologie (CNR-IFN), 20133 Milano, Italy

[4]Department of Physics, University of California, Berkeley, California 94720, USA

[a]Current Address: Department of Chemistry and PULSE Institute, Stanford University, Stanford, CA 94305, USA

---

[*] These authors contributed equally to this work.
[⊥] email: ogessner@lbl.gov

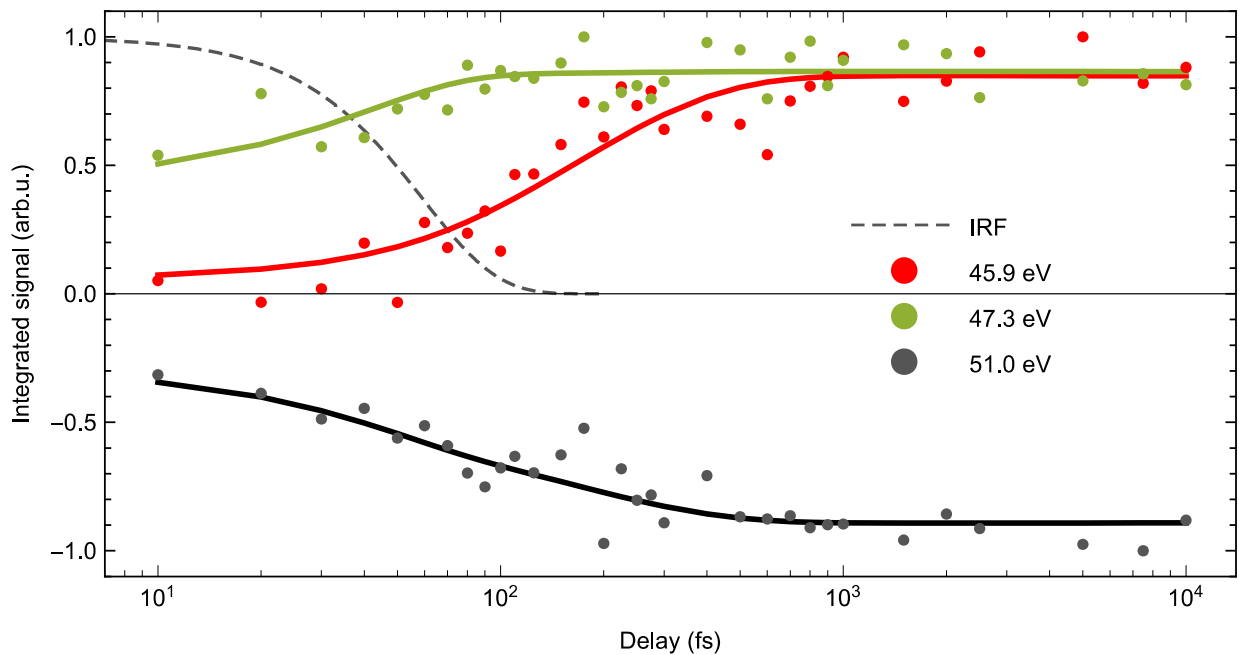

Fig. S1. Time-dependent spectral trends as shown in Fig. 1B of the main text, with the delay plotted on a logarithmic scale and extending to 10 ps.

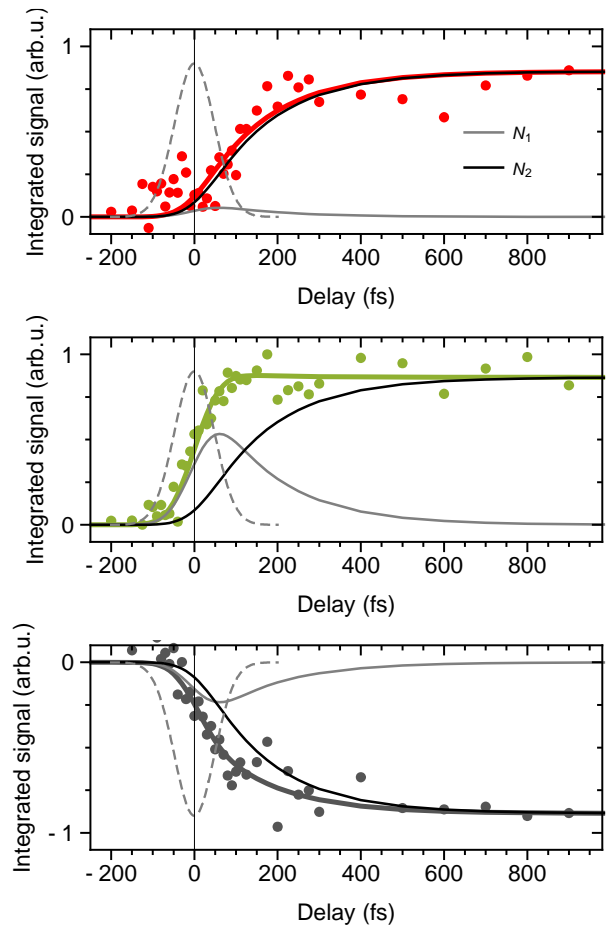

Fig. S2. Time-dependent contributions of $N_1$ and $N_2$ to the overall fit shown in Fig. 1B of the main text.

In the main text we make the approximation that the neutral (I atom) and ionization (I$^+$, I$^{2+}$) product channels can be described by a simple system of differential equations:

$$\frac{dN_1}{dt} = g(t - t_0, \sigma) - k_1 N_1(t) \tag{1}$$

$$\frac{dN_2}{dt} = k_1 N_1(t) \tag{2}$$

However, this intrinsically couples the time-dependent changes of the neutral dissociation and ionization channels, as only the fitting parameter $k_1$ describes population flow from $N_1$ to $N_2$ and must describe the whole dataset. Decoupling the two channels, such that each has an independent 2-step pathway, results in the following system of differential equations:

$$\frac{dN_{1,neu}}{dt} = g(t - t_0, \sigma) - k_{1,neu} N_{1,neu}(t) \tag{3}$$

$$\frac{dN_{2,neu}}{dt} = k_{1,neu} N_{1,neu}(t) \tag{4}$$

$$\frac{dN_{1,ion}}{dt} = g\left(t - t_0, \frac{\sigma}{\sqrt{2}}\right) - k_{1,ion} N_{1,ion}(t) \tag{5}$$

$$\frac{dN_{2,ion}}{dt} = k_{1,ion} N_{1,ion}(t) \tag{6}$$

Such a model is appealing as it allows flexibility in the temporal and spectral fit to treat the processes of neutral dissociation and ionization separately, plus it enables an independent σ to be used in the Gaussian population function for the single-photon and multi-photon pathways. However, the additional degrees of freedom can produce unphysical fit results, for example by assigning large positive and negative terms to the decomposed spectra that, in linear combination, reproduce the dataset (see Fig. S3).

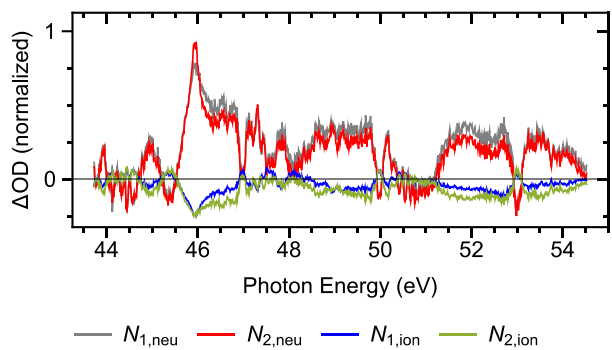

Fig. S3. Failed decomposition from excessive free-fit parameters.

To simplify the fit, we reduce the number of free fit parameters by combining the signals associated with the neutral state $N_{1,neu}$ and the ionic state $N_{1,ion}$ into a single intensity $N_{1,ion}$. This does not mean that no signal from the initially excited neutral molecule is detected, but it is simplification of the fit to avoid artefacts. In reality, both the ion and neutral channels have overlapping contributions. However, they are difficult to separate as they overlap both spectroscopically and on timescales < IRF. The fit then returns decomposed spectra with some key qualities (see Fig. S4): the spectrum of $N_{2,neu}$ contains not just the I atom signal at 45.9 eV, but also the weaker I* transition probed at 46.7 eV. This gives confidence that the spectral contributions are partitioned into $N_{2,neu}$ and $N_{2,ion}$ at the asymptote correctly. An estimate of the branching between the neutral and ion channels can be obtained by inspecting the region of the parent depletion signal; in Fig. S4C (right column) the amplitudes of $N_{2,neu}$ (red) and $N_{2,ion}$ (green) pathways are approximately equal, with substantial uncertainty. The timescales from this neutral plus ionization model are $\tau_{1,neu} = 160 \pm 30$ fs and $\tau_{1,ion} = 15 \pm 15$ fs, in agreement with the simpler estimate for neutral I atom formation of $\tau = 165 \pm 30$ fs in the main text and consistent with ion formation on timescales less than the IRF. This model suggests positive absorption from an intermediate molecular species at 51.8 eV and 53.5 eV, which overlaps with the depletion of the parent molecule.

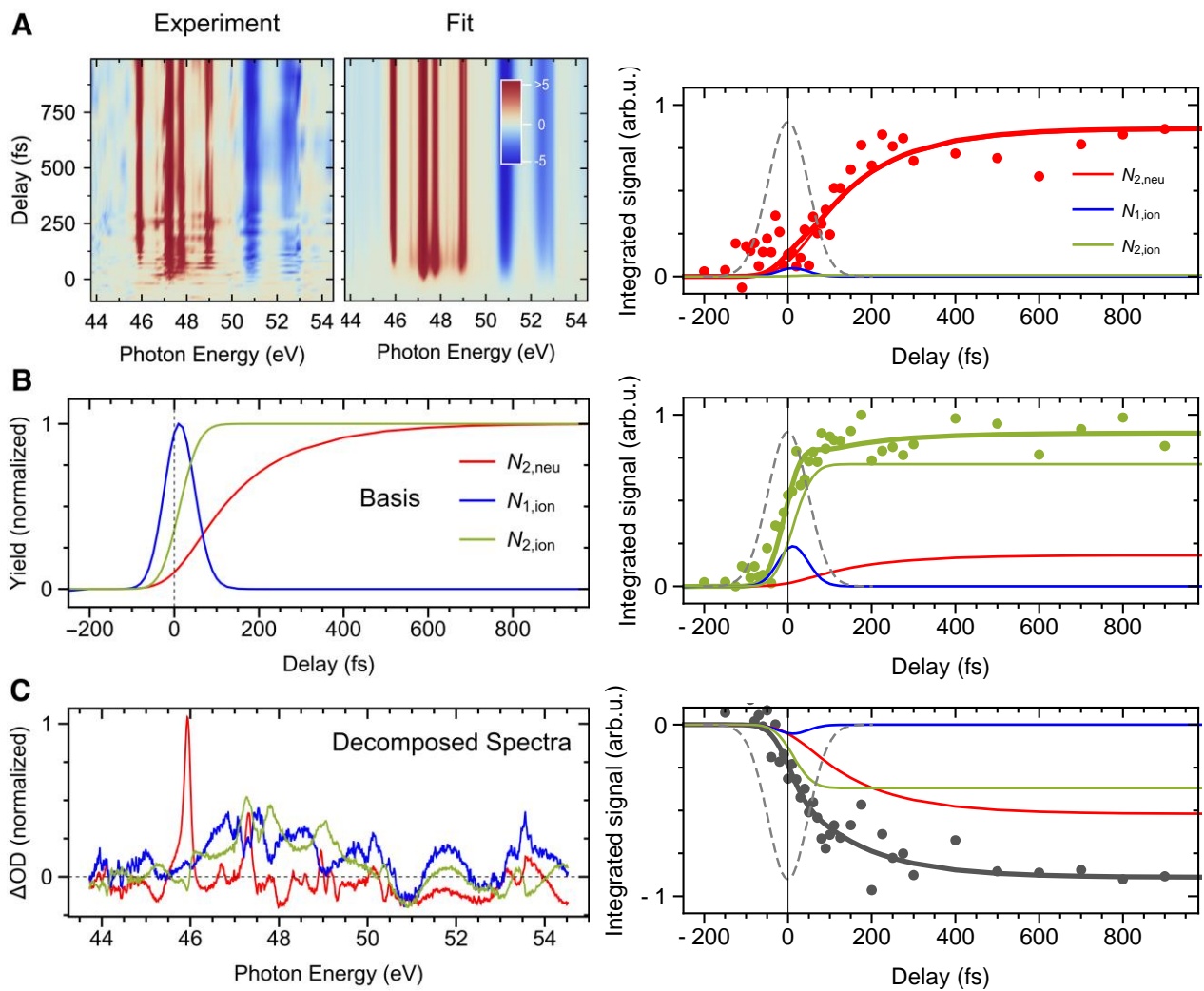

Fig. S4. **Left:** Global fit analysis of the time-resolved XUV absorption spectrum of 2-iodothiophene. **(A)** Measured and global fit data as a function of XUV absorption energy and pump-probe time delay. **(B)** The time basis set is defined by two parallel two-state models where each state is sequentially populated, one for neutral and one for ionic fragments. $N_1$ is populated by a Gaussian population function and depopulated into $N_2$ by a rate constant determined from the fit. **(C)** Decomposed spectra. **Right:** Time-dependent contributions of fit components to the overall fit shown in panel **(A)** at 45.9 eV (red), 47.3 eV (green) and 51.0 eV (black), corresponding to I, I+ and parent depletion.

**Computational details**

Calculations were carried out with Kohn-Sham density functional theory (KS-DFT)[1] using the cam-B3LYP functional[2] and the Q-Chem 5 package.[3] All local exchange–correlation DFT integrals were calculated over a radial grid with 99 points and an angular Lebedev grid with 590 points for all atoms.

| Excited State | TDDFT/TDA [4] | TDDFT | ROKS [5] | EOM-CCSD | EOM-CCSD(fT) |
|---|---|---|---|---|---|
| $\pi \to \sigma^*$ | 4.46 | 4.40 | 4.39 | 4.61 | 4.28 |
| $n_\perp \to \sigma^*$ | 5.31 | 5.23 | 5.24 | 5.43 | 5.23 |
| $\pi \to \pi^*$ | 5.50 | 5.30 | 5.27 | 5.60 | 5.17 |

Table S1: Vertical excitation energies (in eV) of the three lowest valence singlet excited states at the optimized ground state geometry of 2-iodothiophene, with different excited state DFT approaches (using cam-B3LYP/def2-TZVPD[6]). The $\pi \to \pi^*$ excitation is experimentally located at 5.2 eV. EOM-CCSD[7] and EOM-CCSD(fT)[8] results (using the aug-cc-pVTZ[9-11] basis on all atoms, and the associated pseudopotential on I) for the same, DFT optimized, geometry are also provided for comparison.

Ground state and time-dependent DFT (TDDFT) calculations on 2-iodothiophene were performed with the def2-TZVPD basis (with the corresponding pseudopotential for I), for optimizing structures and computing valence excitations. Valence excitation energies obtained with restricted open-shell Kohn-Sham (ROKS) agree very well with TDDFT, but slightly less so with the Tamm-Dancoff approximation or TDA (as shown in Table S1). Furthermore, the TDDFT results agree reasonably well with the EOM-CCSD(fT) approach. All TDDFT calculations therefore were done without TDA, unless noted otherwise.

Fewest switches surface hopping (FSSH)[12] *ab initio* trajectory calculations were also carried out, using TDDFT and the smaller def2-SV(P) basis. The FSSH calculations started on the $\pi \rightarrow \pi^*$ electronic excited state, but with the optimized (cam-B3LYP/def2-SV(P)) ground state geometry. The initial nuclear velocities for the FSSH trajectories were obtained by supplying each ground state normal mode with kinetic energy equal to the ground state zero-point energy for the said mode (while the sign of the mode velocity was selected randomly). The total kinetic energy thus supplied in this quasiclassical (QC)[13] manner is 36 kcal/mol (1.56 eV), which corresponds to the total ground state vibrational zero-point energy (at the cam-B3LYP/def2-SV(P) level). The FSSH trajectories were run until the lowest TDDFT singlet excitation energy became imaginary[1], which indicates that a bond has been essentially completely broken. The exact duration of an individual trajectory is thus a function of the specific dynamics in that trajectory, but this approach ensures that all trajectories exhibited a bond dissociation event. In practice, the typical trajectory duration was ~100 fs (with individual timesteps of 0.48 fs duration). FSSH statistics were collected over 100 trajectories, with C-I dissociation occurring in 67 of those (and ring opening via C-S cleavage in the rest). Bootstrap sampling indicates that the one standard deviation uncertainty, arising from the finite number of initial nuclear velocities (100) utilized, is 5%.

A histogram of C-I dissociation times over the 67 trajectories is given in Fig. S5. The augmented FSSH approach[14] (which adds decoherence) yielded similar results, as did FSSH with a different functional (PBE0[15]), indicating that the results are not too sensitive to these aspects of the calculation. Spin-unrestricted TDDFT was not attempted due to discontinuities in the energy surface that arise in calculations.[16]

---

[1] This indicates that a bond dissociation has already happened long before, as the system must be stretched considerably beyond the Coulson-Fischer point (which marks the onset of spin-polarization) for the singlet TDDFT excitation energies to cease to be real.

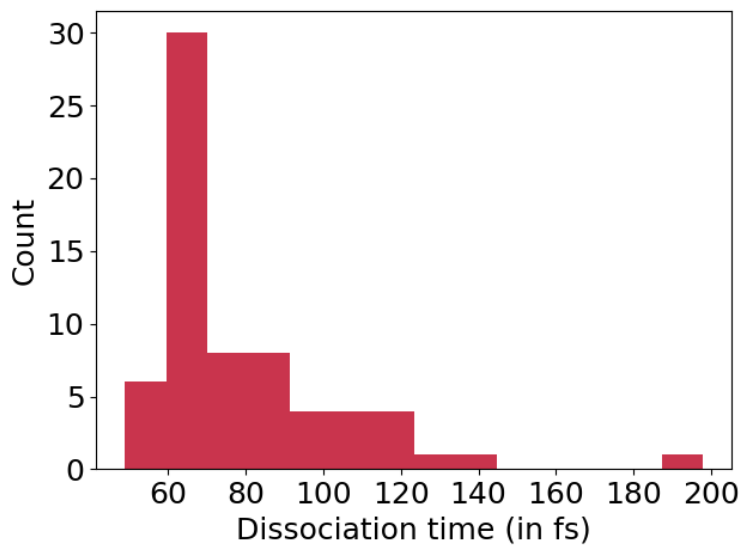

Figure S5. Histogram of times taken by individual FSSH trajectories to reach a C-I separation of 270 pm. Only results for the 67 (out of 100) trajectories with C-I dissociation are shown.

The singlet state of ring-opened 2-iodothiophene is biradical in character, and thus it is not readily accessed via standard KS-DFT. ROKS was used to optimize the geometry of this state instead (with cam-B3LYP/def2-TZVPPD). However, the corresponding triplet can be characterized by KS-DFT with spin-unrestricted orbitals without issues. QC trajectory calculations on this triplet (starting from the equilibrium geometry) were performed at the cam-B3LYP/def2-SV(P) level to estimate the rate of "hot" I atom ejection from the open-ring product. 1.5 eV excess kinetic energy (corresponding to the energy difference between the open-ring triplet structure and ground state 2-iodothiophene pumped by 4.6 eV extra energy) was initially supplied to the molecule by a random distribution over normal modes, on top of the zero-point energy of the open-ring triplet structure (1.46 eV). These trajectories were run for 1.8 ps (with timestep duration being 1.2 fs).

The iodine N-edge calculations used aug-cc-pωCVTZ[17] for I (along with the associated pseudopotential) and aug-cc-pVDZ on all other atoms. Core-excited states for neutral species were modeled with ROKS, while the reference state was computed with either standard KS-DFT (for the closed-shell 2-iodothiophene ground state) or ROKS (for biradical starting states like 2-iodothiophene singlet excited states or the singlet open-ring species). The first core-excited state of the cation (corresponding to SOMO←4d excitation) was modeled with ΔSCF,[18] as only one unpaired electron was present.

All the geometries employed in this work have been supplied in the supplementary material (*zip).

**Choice of density functional**

cam-B3LYP was chosen as the functional for two reasons. It is widely used for valence TDDFT calculations due to the presence of range-separated HF exchange, and indeed yields a reasonable reproduction of the features in the experimental gas-phase UV-Vis absorption spectrum of iodothiophene (as can be seen in Fig S6.). Furthermore, vertical TDDFT excitation energies from cam-B3LYP agree reasonably well with EOM-CCSD(fT), as shown in Table S1.

cam-B3LYP also has a distinct advantage of being quite accurate for orbital-optimized core-excitation energy computations, attaining 0.2-0.3 eV scale errors for K-edges of C, N, O, and F. We find that it is likely to be reliable for the iodine N-edge calculations as well. Specifically, it yielded 50.5 eV for the lowest N-edge excitation of $CH_3I$ (vs 50.6 eV from experiment), 45.5 eV for a single I atom (vs 45.9 eV from experiment) and most importantly, 50.8 eV for 2-iodothiophene (vs 50.9 eV from experiment). Other reasonable functionals for core-excitations (like SCAN[19] or $\omega$B97X-V[20]) yielded similar iodine N-edge values, indicating that orbital-optimized DFT is likely as reliable for iodine N-edge calculations as it has been shown to be for the K-edge of light elements. However, these functionals are somewhat poorer at predicting the valence excitations of 2-iodothiophene (as can be seen from Fig S6), making cam-B3LYP the optimal choice for both iodine valence and N-edge computations.

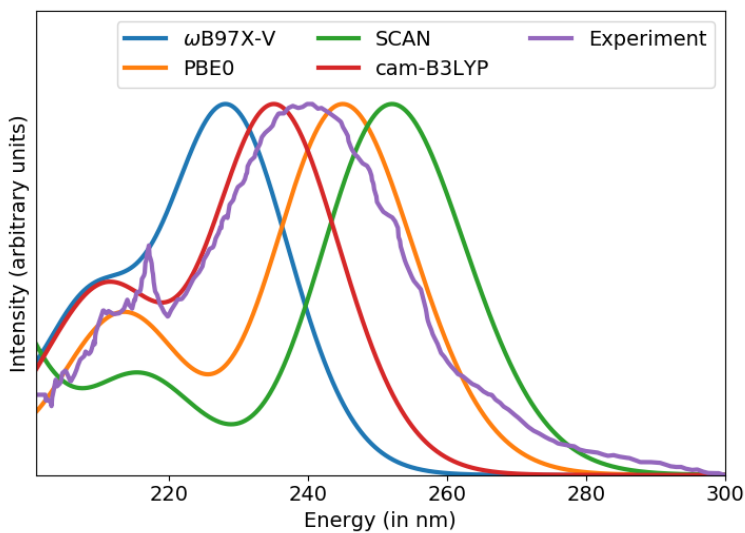

Figure S6. Gas-phase TDDFT UV-vis absorption spectra of 2-iodothiophene predicted by some DFT functionals, as compared to experiment. Use of ROKS instead of TDDFT yields a similar picture.

**Calibration of iodine N-edge computed spectra**

One interesting factor to consider is that the calculations (which are non-relativistic) yielded the energy of the lowest multiplet without any spin-orbit correction (the 4d multiplet splitting being 1.6 eV for I). This is likely just a consequence of the pseudopotential employed (as different pseudopotentials yield somewhat different results, as can be seen from Table S2), which is not really optimized for iodine N-edge computations and thus cannot really be expected to yield the multiplet averaged result. The consistency of different functionals for a given pseudopotential on the other hand (as shown in Table 2), indicates that ROKS is yielding a very good estimate for the exact excitation energy for any given pseudopotential. The case of $CH_3I$ shows that the results are fortuitously close to the lowest multiplet energy for the aug-cc-pωCVTZ basis and associated pseudopotential. We therefore report the *spin orbit* *uncorrected* results for this pseudopotential as the lower energy multiplet for any given excitation (and find the higher energy multiplet by adding 1.6 eV to this number).

| Basis set | cam-B3LYP | SCAN | ωB97X-V |
| --- | ---: | ---: | ---: |
| aug-cc-pωCVTZ/aug-cc-pVDZ | 50.53 | 50.60 | 50.67 |
| unc def2-TZVPPD/def2-SVPD | 50.57 | 50.62 | 50.71 |
| unc 6-311G*/6-31+G* [21] | 53.56 | 53.75 | 53.68 |
| CRENBL[22,23] | 56.62 | 56.76 | 56.92 |

Table S2. Lowest iodine N-edge excitation energy (in eV) of CH$_3$I with various functionals and pseudopotentials/basis sets. The experimental reference value is 50.6 eV. A basis reported as X/Y indicates use of X basis functions for I and Y basis functions for the C and H atoms. For example, "unc *def*2-TZVPPD/def2-SVPD" means the uncontracted ("unc") *def*2-TZVPPD basis was used for I and def2-SVPD for other atoms.

**Contributions from multi-photon ionization**

The transient XUV spectra in Fig. 1 and Fig. 2 exhibit significant contributions from atomic iodine ions to the product distribution. Thus, ionization by absorption of multiple 268 nm photons cannot be neglected and the question arises to what extent dissociative multi-photon ionization may contribute to the neutral iodine signal. To this extent, Fig. S7 shows a comparison of XUV spectra for two different pump pulse intensities of 1 TW/cm$^2$ (black) and 5 TW/cm$^2$ (gray). As expected, the production of ionic species is greatly enhanced at the higher intensity. Notably, however, the signal intensity from neutral iodine fragments in the range ~45 eV - 47 eV remains virtually unchanged, indicating that multi-photon ionization does not contribute to a significant amount to the neutral iodine product channel.

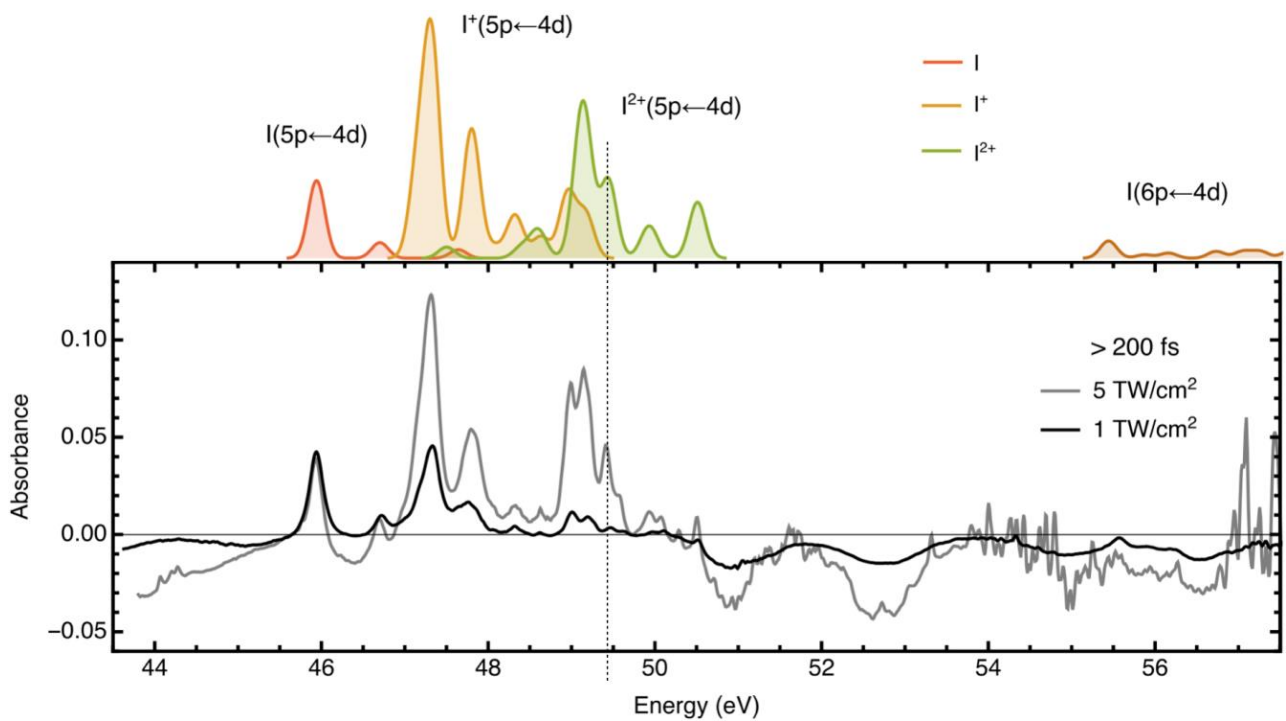

Figure S7. Dependence of XUV absorption spectra at long pump-probe delays >200 fs on pump pulse intensity. While ionic signal contributions are greatly enhanced at 5 TW/cm$^2$ (gray) compared to 1 TW/cm$^2$ (black), the intensities of neutral iodine signals in the range ~45 eV - 47 eV remain virtually unchanged, indicating that ionic and neutral iodine products emerge from independent processes. The vertical dashed line marks the energy where the strongest isolated I$^{2+}$ signal is expected.

While the 5 TW/cm$^2$ data in Fig. S7 show indications for contributions from doubly charged I$^{2+}$ fragments (vertical dashed line), this product does not appear in appreciable amounts in the 3 TW/cm$^2$ measurements presented in the main text, as illustrated in Fig. S8.

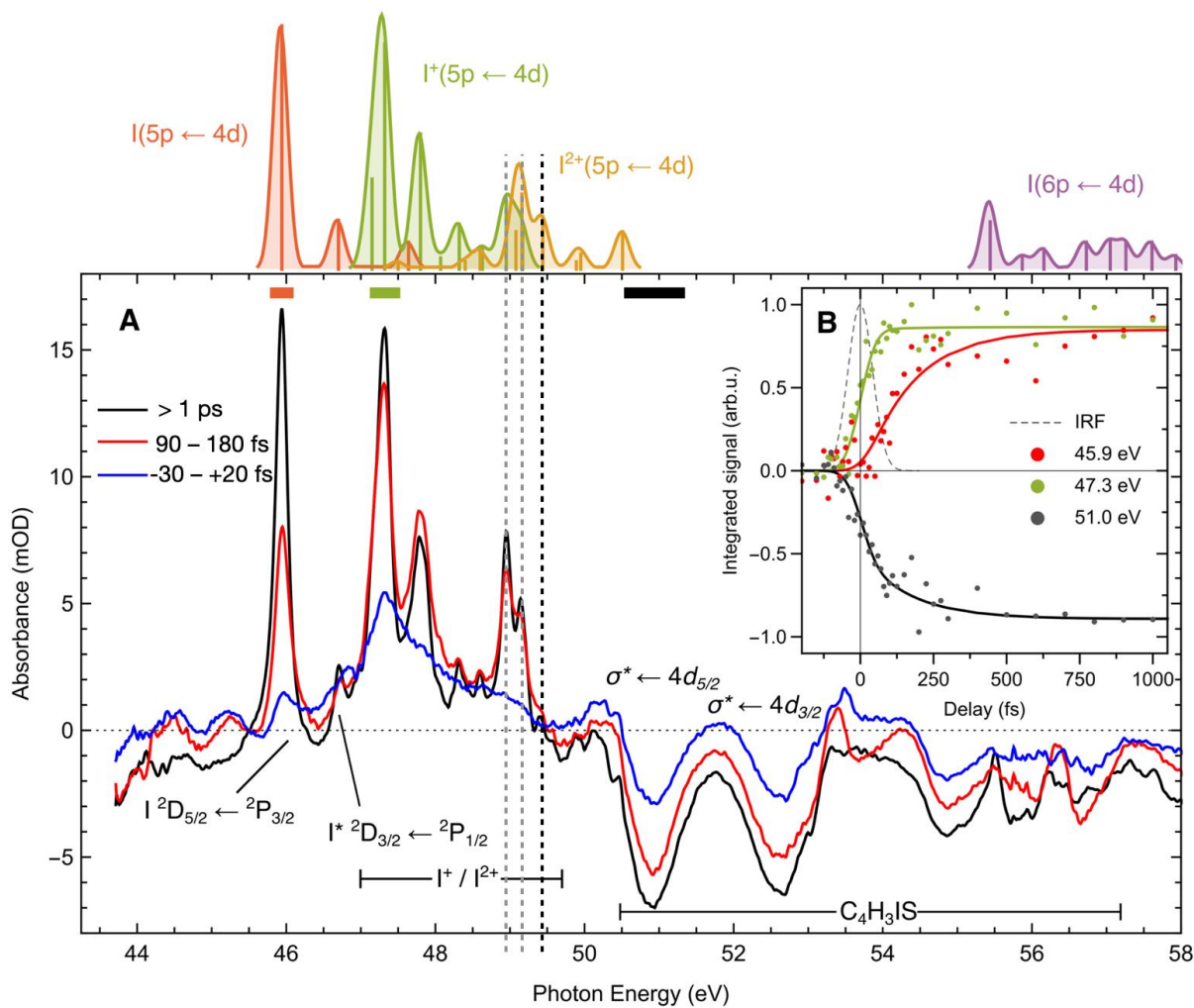

Figure S8. Contributions from doubly charged $I^{2+}$ products appear minor at the pump pulse intensity of 3 TW/cm² used in the measurements presented in the main text. The gray vertical dashed lines indicate the highest energies of the $I^+$ spectrum and the black dashed line the location of the strongest isolated $I^{2+}$ signal.

Note that comparison of the absolute absorbance values is challenging since there can be small changes in experimental conditions on different days, in particular regarding the sample density, which can have notable impact on these absolute numbers. The relative magnitudes, however, of neutral and charged species signals are largely unaffected by these changes. Thus, the relative signal trends indicate that the yields of neutral and ionic fragments show distinct behavior with change in the pump laser power, with the ionic product yield rising much faster with increasing power, as expected from the multi-photon processes required to ionize the molecule. We also note that the appearance times of the neutral iodine fragments are comparable for measurements made at the three different pump powers, supporting the notion that the underlying dynamics are not dominated by multi-photon effects.

As mentioned in the main text, we reemphasize that parent molecular ion signals are only expected above 48.0 eV photon energy. Thus, the dominant peak at early times, located at ~47.3 eV, can safely be assigned to neutral species and, with the aid of the *ab initio* results presented in Table 1, to the initially populated π→π* excited state.

**Ring-opening processes**

Based on the results of their ion imaging study and, in particular, high-level *ab initio* calculations, Marchetti and co-workers point out a distinct possibility for ring-opening in UV-excited 2-iodothiophene as a consequence of nonadiabatic coupling between initially accessed π→π* states and (n/π) →σ$_{C-S}$* excited states that are unstable with respect to C-S bond extension.[24] Complementary time-resolved infrared absorption and XUV photoemission studies on the structurally similar 2(5H)-thiophenone ($C_4H_4OS$) molecule revealed clear evidence for very rapid (<100 fs) emergence of ring-open configurations in $S_2$ and $S_1$ electronically excited states after ~265-267 nm excitation.[25,26] Subsequent relaxation to vibrationally hot $S_0$ ground state molecules was determined to proceed on a timescale of ~350 fs. Natural difference orbitals (NDO) of the Franck-Condon geometry show that the initial valence-excited $S_2$ electronic state is primarily of n$_S$→π* in character, where excitation occurs from a nonbonding *p* orbital on the sulfur atom into a π* orbital of the ring. A smaller, yet non-negligible contribution from a n$_O$→σ* excitation involving an antibonding σ* orbital along the S-C(-O) bond and a minor contribution from a π→π* transition have also been identified. Trajectory surface hopping calculations suggest the initial nuclear motions involve C–S bond extension, to reach a series of conical intersections and ultimately funnel population to $S_0$. Once on $S_0$, the trajectories show interconversion between several ring-opened isomers, even transiently ring-closing to reform thiophenone, with some ultimately undergoing fragmentation to form thioacrolein + CO fragments over tens to hundreds of picoseconds. Our calculations also indicate the possibility of C-S bond cleavage for 268 nm excited 2-iodothiophene (albeit without the simultaneous breaking of the C-I bond) with ~1/3 of the FSSH *ab initio* trajectories resulting in ring-open products. Notably, however, no indication for any significant contribution of such products to the XUV absorption spectra has been detected, which would be expected at 49.3 eV (f ~ 0.0025) and 49.8 eV (f ~0.020). In particular, the higher energy transition has significant oscillator strength and is located in a relatively uncongested spectral region, where strong intermediate and/or final product signals could be readily distinguished. A potential source of the discrepancy arises from the use of the $S_0$ geometry as the initial configuration in the AIMD calculations, which is a better approximation to a pump excitation corresponding to the band maximum (~240 nm) as opposed to the 268 nm UV wavelength employed in the experiment. This therefore suggests that use of shorter pump wavelengths could lead to a greater proportion of ring-opened products. Considering the significance of these channels for (photo)chemical transformations, an important task for future studies is to further investigate the wavelength dependence of different channels in this respect.